# Electrostatic control of valley-dependent phase in tilted Dirac/Weyl channels

Can Yesilyurt[a)]

*Nanoelectronics Research Center, Istanbul, Turkey*

Valley degrees of freedom represent a promising resource for solid-state quantum information. This work demonstrates electrostatic control of the valley-dependent phase in tilted Dirac/Weyl semimetals. When wave packets traverse a shaped electrostatic barrier, the valley-dependent tilt induces differential spatial drift and dwell times, resulting in the accumulation of a continuously tunable relative dynamical phase. The barrier thereby functions as a valley-phase shift element. Time-dependent transport simulations employing a coupled two-valley model yield electrically tunable relative phases between equal-energy valleys, including $\pi/4$, $\pi/2$, and $\pi$, with high transmission probabilities.

[a)]Author to whom correspondence should be addressed: cyesil@nanorescenter.com; can--yesilyurt@hotmail.com

## I. INTRODUCTION

Within solid-state quantum architectures, electron-based qubits are especially attractive because they offer fast electrical tunability and compatibility with state-of-the-art nanofabrication. Historically, the reference point for semiconductor qubits has been the electron spin in coupled quantum dots, supported by a vast body of work on spin initialization, coherent control, and exchange-based gating.[1–3]

When a material's band structure contains inequivalent valleys in momentum space, an additional binary quantum number becomes available. Initially viewed merely as a constraint on silicon spin-qubit design, this valley degree of freedom has matured into a useful computational resource. Theoretical proposals have highlighted valley encoding as a route to noise-resilient quantum computation,[4] and experiments have since demonstrated the coherent manipulation of silicon valley states, electrically tunable valley splitting, and subnanosecond two-axis control.[5–7] Recent observations of long-lived valley states in bilayer graphene quantum dots further strengthen the case for valley quantum numbers as robust information carriers.[8]

However, the vast majority of valley-qubit architectures exploit confinement- or interface-induced valley energy splitting as the primary control mechanism.[3,5,6] This paradigm does not easily transfer to the emerging class of Dirac and Weyl materials. In these related topological systems, quasiparticles are effectively massless, their velocity is governed by a linear energy dispersion rather than a large effective mass, and the valleys (or nodes) are generally separated in momentum space while remaining protected by symmetry.[9–11] Crucially, symmetry-related valleys or nodes can remain exactly degenerate in energy even while being separated in momentum space: in time-reversal-symmetric tilted Dirac/Weyl systems, the cone tilt reverses sign between partner valleys. Manipulating these states, therefore, presents a distinct challenge: one must achieve coherent control of the valley degree of freedom without relying on built-in valley energy splitting.

Transport through Dirac barriers offers rich, valley-dependent phenomenology. Prior studies have established valley filtering and valley valves in graphene,[12–15] near-perfect transparency via Klein tunneling,[16] and valley-contrasting Berry-curvature responses.[17] In tilted Dirac/Weyl systems, transport becomes even more distinctive, as the energy tilt steers opposite valleys along different refraction pathways, enabling electron optics and classical valley polarization.[18,19] This literature, however, operates on valley *populations*: valleys are filtered, polarized, or split by breaking valley symmetry with magnetic fields, strain, or magnetoelectric barriers, or by exploiting p–n junctions and Klein-tunneling transmission asymmetries, and the observables are transmission intensities and classical trajectories. Moreover, the Fabry–Pérot phases of p–n structures are common to both valleys and are therefore valley-independent. To our knowledge, the coherent *phase difference* between valleys—the quantity that turns propagating valley states into a quantum resource—has not been reported. In the present scheme, a purely electrostatic barrier drives order-$\pi$ action in the phase channel while the intensity channel remains deliberately silent.

A shaped electrostatic barrier within a confined, tilted Dirac/Weyl channel can function as a coherent, electrically tunable valley-phase-shift element. The underlying mechanism is as follows: valleys with different chiral structures at the same energy level accumulate distinct dynamical phases during coherent propagation. Because the Dirac cone tilt reverses sign between valleys, the wave packets experience differential transverse drift during transport. In the low-barrier regime, the structure primarily introduces a controllable relative phase rather than acting as a valley filter. This approach provides a transport-based route to coherent manipulation of Dirac/Weyl fermions, thereby enabling quantum information processing with relativistic valley qubits.

## II. RESULTS

### A. Electrostatically controlled valley-phase shift element in a tilted Dirac/Weyl channel

We describe coherent transport in the channel by a four-component two-valley model in which the spinors of the two valleys form a single state $\Psi = (\psi_K, \psi_{K'})$ that evolves under

$$H = \begin{pmatrix} H_K & U_{KK'} \\ U_{KK'}^{\dagger} & H_{K'} \end{pmatrix}, \tag{1}$$

$$H_\tau = \tau\,\hbar\,\mathbf{w}\cdot\mathbf{k}\,\mathbb{I} + \hbar v_{\mathrm{F}}\big(\sigma_x k_x + \tau\,\sigma_y k_y\big) + V(x,y)\,\mathbb{I} + M(y)\sigma_z, \tag{2}$$

where the valley index $\tau = +1$ $(K)$ and $-1$ $(K')$ sets both the tilt sign and the chirality, $\mathbf{k} = (k_x, k_y)$ is the wavevector, $v_{\mathrm{F}}$ is the Fermi velocity, $\mathbf{w} = (w_x, w_y)$ is the tilt vector, $V(x,y)$ is the shaped electrostatic barrier, $M(y)\sigma_z$ provides the reflecting sidewall confinement, and $U_{KK'}(x,y)$ is an intervalley coupling field. Time reversal enforces both the tilt reversal and the conjugate chirality of the $K'$ block, $H_{K'}(\mathbf{k}) = H_K^*(-\mathbf{k})$. Smooth potentials carry no Fourier weight at the valley separation $|K-K'|$, so the engineered gate and confinement elements do not couple the valleys; $U_{KK'}$ represents atomic-scale perturbations, and the robustness of the phase operation against it is quantified in Supplementary Note 1. Considering an ideally clean device channel, the results below correspond to the limit $U_{KK'} \to 0$, in which the coupled propagation reduces to exactly two independent valley evolutions (see Appendix 4).

In this limit, the barrier cannot transfer amplitude between the two momentum-separated valleys, which form a natural pseudospin basis:

$$|0\rangle_v \equiv |K\rangle, \qquad |1\rangle_v \equiv |K'\rangle, \tag{3}$$

and the entire coherent action of the barrier within this subspace is a relative phase. We define it relative to the corresponding zero-barrier evolution at a common postscattering time $t^*$. Because bare tilted propagation already separates the two valleys transversely, the relevant comparison basis at that plane is the pair of transported reference modes $\{|\Psi_K^{(0)}(t^*)\rangle, |\Psi_{K'}^{(0)}(t^*)\rangle\}$ rather than a common drain-local orbital mode. For each valley, the barrier–reference overlap is

$$\mathcal{O}_\tau(V_0;t^*) = \langle \Psi_\tau^{(0)}(t^*) \mid \Psi_\tau^{(V_0)}(t^*)\rangle = |\mathcal{O}_\tau| e^{i\phi_\tau}, \tag{4}$$

where $\Psi_\tau^{(0)}$ and $\Psi_\tau^{(V_0)}$ are the zero-barrier and barrier-propagated spinors, respectively. The barrier-induced relative phase is then

$$\Delta\phi = \arg[\mathcal{O}_K\, \mathcal{O}_{K'}^*] = \phi_K - \phi_{K'}, \tag{5}$$

and each barrier-propagated valley can be decomposed as

$$|\Psi_\tau^{(V_0)}(t^*)\rangle = \mathcal{O}_\tau\, |\Psi_\tau^{(0)}(t^*)\rangle + |\delta\Psi_\tau^{\perp}(t^*)\rangle, \qquad \langle \Psi_\tau^{(0)}(t^*) \mid \delta\Psi_\tau^{\perp}(t^*)\rangle = 0. \tag{6}$$

For normalized comparison wave functions, $\| \delta\Psi_\tau^{\perp} \| = \sqrt{1-|\mathcal{O}_\tau|^2}$; we refer to $|\mathcal{O}_\tau|^2$ as the retained-mode probability and to $1-|\mathcal{O}_\tau|^2$ as the reference-mode leakage. Within the transported reference subspace, the barrier-induced relative operation is therefore approximately

$$U_{\mathrm{rel}}^{\mathrm{eff}}(t^*) \approx \begin{pmatrix} \mathcal{O}_K & 0 \\ 0 & \mathcal{O}_{K'} \end{pmatrix}. \tag{7}$$

When $|\mathcal{O}_K|$ and $|\mathcal{O}_{K'}|$ are both close to unity, the barrier acts as a nearly pure valley-phase shift element with a relative phase $\Delta\phi$.

The device concept is illustrated in Fig. 1. A single-mode wave packet is injected into a two-dimensional channel bounded laterally by mass walls and traverses a shaped electrostatic barrier featuring a vertical entrance interface and an oblique exit interface. The bare Dirac cone tilt gives the two valleys opposite transverse group-velocity components ($v_{g,y} = \pm w_y$), causing their centroids to propagate on opposite sides of the channel axis. Crucially, the oblique exit interface refracts both valleys in the same transverse direction—toward its normal but by different amounts—because the tilt-reversed valleys arrive at the interface with different effective incident angles and because momentum conservation at the interface couples to the opposite-sign tilt terms differently. This produces a valley-dependent exit position along the oblique interface, thereby resulting in a different effective barrier width for each valley.

The mechanism is intrinsically unidirectional: for normal incidence, the vertical entrance interface introduces no transverse deflection, whereas the oblique exit interface breaks this symmetry. This contrasts with conventional straight barriers, where the two interfaces are parallel and perpendicular to the transport direction.

The barrier region is defined by two linear interfaces:

$$\begin{aligned} x_L(y) &= x_{L,b} + \frac{y + L_y/2}{L_y}(x_{L,t} - x_{L,b}), \\ x_R(y) &= x_{R,b} + \frac{y + L_y/2}{L_y}(x_{R,t} - x_{R,b}), \end{aligned} \tag{8}$$

yielding an oblique exit. For normal injection, the outside-barrier group velocity is valley-dependent because of the tilt, $\mathbf{v}_{g,\tau} = \tau\mathbf{w} + v_F\hat{\mathbf{x}}$. In the low-barrier limit, the resulting phase shift is estimated as follows:

$$\Delta\phi \approx \frac{V_0}{\hbar v_F}\Delta D_{\text{eff}}, \qquad \Delta D_{\text{eff}} \equiv D_{K'} - D_K, \tag{9}$$

where $\Delta D_{\text{eff}}$ is the difference in the valley-dependent effective barrier width. Throughout, the simulations use $E_F = 80$ meV, $v_F = 10^6$ m s$^{-1}$, and the tilt parameter $\zeta_y = w_y/v_F = 0.35$.

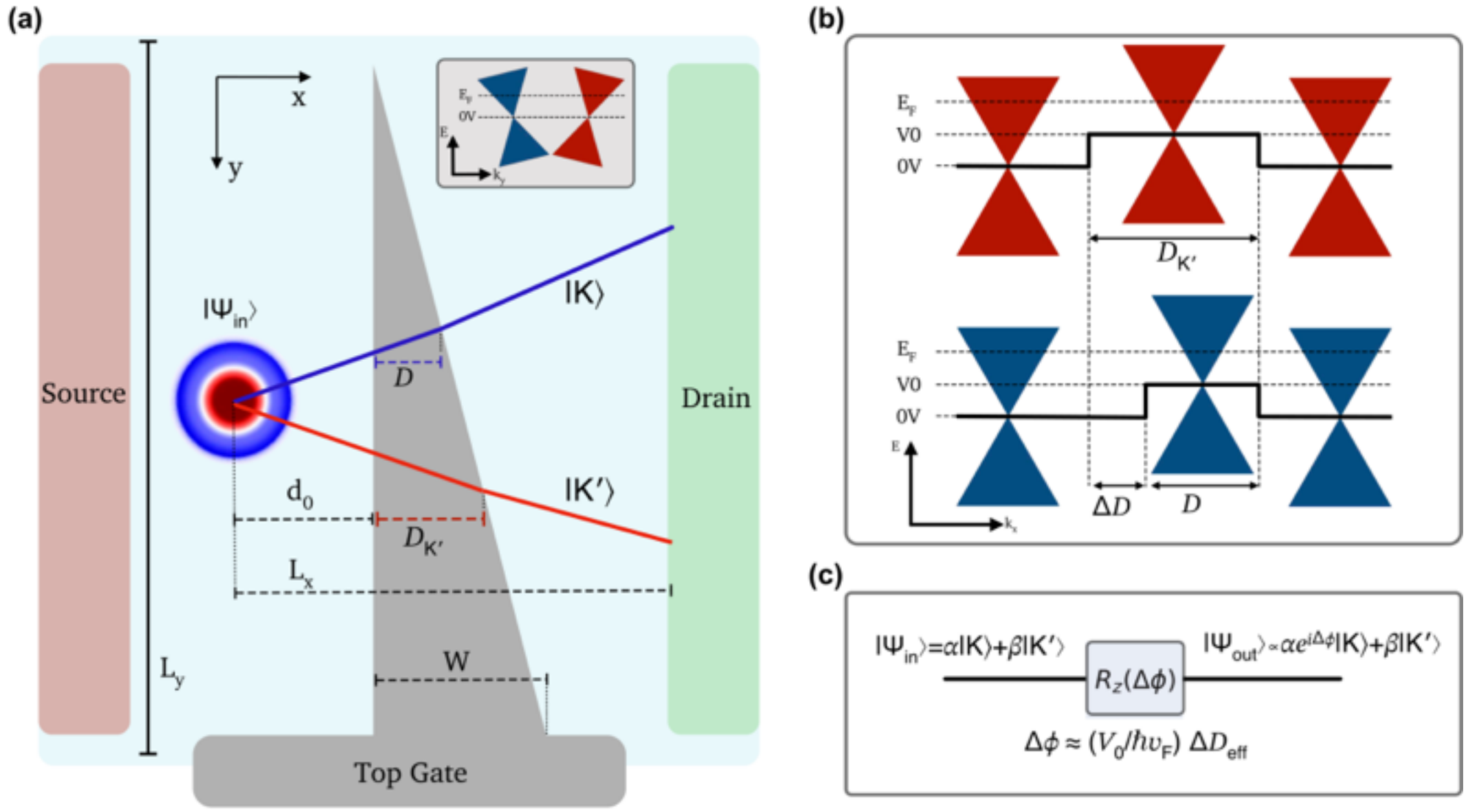

FIG. 1. **Device concept of electrostatic valley-phase control.** **a**, A wave packet $|\Psi_{\text{in}}\rangle$ enters a 2D channel, traversing a shaped electrostatic barrier with vertical entrance and oblique exit. At the $\pi$ point, valley-dependent group velocities and asymmetric refraction cause centroid paths to remain on opposite sides and experience different effective barrier widths $D_K$ and $D_{K'}$. Inset: The barrier shifts the Dirac point while preserving the cone tilt. **b**, Energy diagrams showing Dirac cones along the transport direction before, within, and after the barrier. $K'$ (red) traverses width $D_{K'}$, and $K$ (blue) traverses $D_K$, with $D_{K'} = D_K + \Delta D_{\text{eff}}$. Opposite tilts yield valley-dependent dwell lengths. **c**, The dynamical phase $\Delta\phi \approx (V_0/\hbar v_{\text{F}})\,\Delta D_{\text{eff}}$ induces a relative valley-phase shift, mapping $|\psi_{\text{in}}\rangle = \alpha|K\rangle + \beta|K'\rangle$ to $|\psi_{\text{out}}\rangle \propto \alpha e^{i\Delta\phi}|K\rangle + \beta|K'\rangle$.

## B. Linear control of valley phase difference by applied potential

We systematically examine how both the phase and transport characteristics depend on the applied barrier height, as shown in Fig. 2. In the regime of low barrier height, where $V_0$ is much less than $E_{\text{F}}$, both valleys display nearly perfect transmission with highly symmetric probabilities. This regime thus constitutes the essential operating condition for a phase element, rather than a valley filter, since the valley-transmission asymmetry $|T_K - T_{K'}|$ remains below $10^{-3}$ across the three operating points introduced below, and the valley polarization $\eta = (T_K - T_{K'})/(T_K + T_{K'})$ remains below $5 \times 10^{-4}$. As the barrier height $V_0$ increases toward $E_{\text{F}}$, a mismatch between the Fermi surfaces of the incident and barrier regions induces a crossover to dominant reflection. At the critical point $V_0 = E_{\text{F}}$, the transmission of both valleys drops to roughly one-half and a small valley polarization appears (see Fig. 2a). For barrier heights exceeding $E_\mathrm{F}$, the onset of Klein tunneling in the bipolar regime restores the transmission to near-unity values.

In Fig. 2b, we present the numerically evaluated $\Delta\phi$ together with an analytical approximation based on a rectangular tilted-Dirac barrier, where the effective widths are determined by the device geometry. In the low-barrier regime, the numerical and analytical results are in near-perfect agreement: the phase grows linearly with barrier height at the rate set by the closed-form dwell asymmetry of Eq. (D1), and a full $\pi$ rotation is reached at $V_\pi \approx 0.3\,E_\mathrm{F}$—corresponding to meV-scale barrier heights for the parameters used here.

The dependence of the overlap magnitudes on the barrier height provides direct insight into the mechanisms that limit ideal device behavior. The unwrapped per-valley phases remain smooth and well defined up to relatively high barrier heights (see Fig. 2d), and the overlap magnitudes $|\mathcal{O}_K|$ and $|\mathcal{O}_{K'}|$ remain close to unity throughout the intended operating window (see Fig. 2c). As the barrier height increases beyond this window, the reduction in overlap arises from two distinct physical effects. First, in the range between $25\,\mathrm{meV}$ and $E_\mathrm{F}$, increasing reflection at the barrier leads to a decrease in the full-wave-function overlap. Second, for $E_\mathrm{F} < V_0 < 2E_\mathrm{F}$, the system enters the bipolar Klein regime, where negative refraction at both barrier interfaces causes the transmitted trajectory to bend away from the zero-barrier reference at both the entry and the exit. This results in an effective doubling of the transverse deflection relative to the reference, which further reduces the overlap. Compared with the $K$ valley, the $K'$ valley consistently exhibits a faster decay of overlap even though the $K$ packet shows a larger absolute transverse centroid shift: the overlap is dominated by the momentum-space (deflection) offset rather than the position offset, and $K'$ has a greater deflection (see Supplementary Note 2). Consequently, the optimal operating window is confined to the low-barrier regime, where high transmission, matched amplitudes, and an electrically tunable relative phase can be simultaneously achieved.

The intervalley coherence $\mu = \langle \chi_{K'} | \chi_K \rangle$ (shown in Fig. 2c) is evaluated between the transmitted, individually normalized valley states after centroid alignment (see Appendix 6). This quantity involves no reference evolution and directly bounds the visibility of a valley-interference experiment: $|\mu| > 0.92$ across the operating window, and its phase reproduces the reference-based $\Delta\phi$ to within 1% of $\pi$ at all three operating points. The gap between $|\mu|$ and $|\mathcal{O}_\tau|$ is informative: the common-mode part of the barrier-induced deflection cancels between the valleys, so the cross-valley leakage $1 - |\mu|^2$ remains less than half the reference-mode leakage; Supplementary Note 2 decomposes the latter and shows that $\sim 98\%$ of it is a rigid, coherent phase-space displacement.

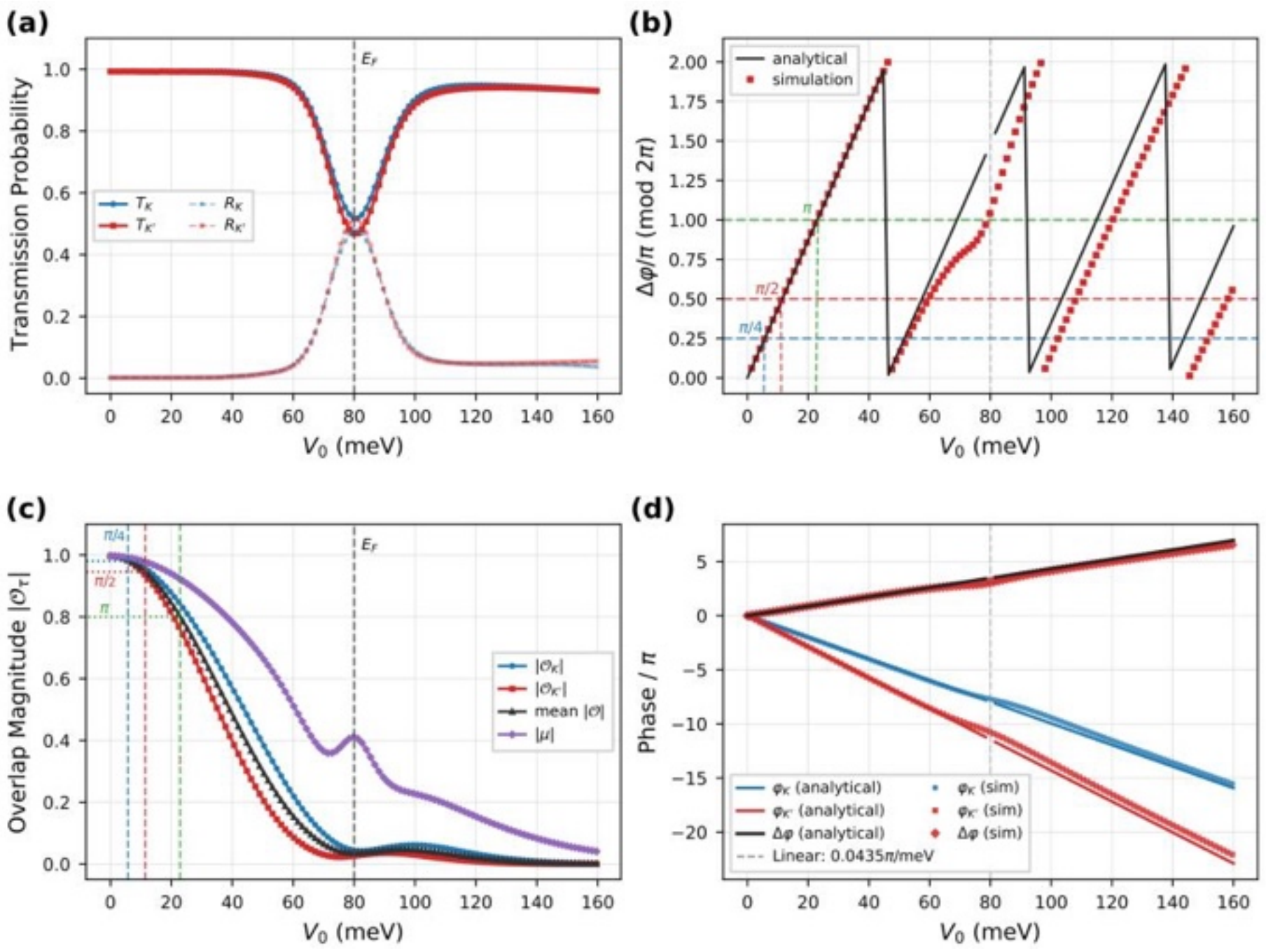

FIG. 2. **Barrier-height dependence of transport, relative phase, and overlap quality. a**, Valley-resolved transmission ($T_K$, $T_{K'}$) and reflection ($R_K$, $R_{K'}$) probabilities as functions of barrier height $V_0$. At low $V_0$, both valleys exhibit nearly perfect transmission and minimal asymmetry, consistent with phase-element behavior. As $V_0$ approaches $E_{\mathrm{F}}$, reflection becomes dominant because of the Fermi-surface mismatch. Above $E_{\mathrm{F}}$, Klein tunneling restores high transmission. **b**, Relative phase $\Delta\phi/\pi$ (mod $2\pi$) with an analytical approximation. Dashed lines indicate $\pi/4$, $\pi/2$, and $\pi$ targets with corresponding barrier heights. The simulation and analytical results agree well, diverging only near $E_{\mathrm{F}}$ because of finite-size effects. **c**, Overlap magnitudes $|\mathcal{O}_K|$ and $|\mathcal{O}_{K'}|$, their mean $\bar{F} = (|\mathcal{O}_K| + |\mathcal{O}_{K'}|)/2$, and the directly calculated intervalley coherence $|\mu|$ as functions of barrier height. Dashed lines mark values at the $\pi/4$, $\pi/2$, and $\pi$ operating points. $K'$ decays faster than $K$ because it has greater momentum-space deflection (Supplementary Note 2). **d**, Unwrapped per-valley phases $\phi_K$, $\phi_{K'}$, and their difference as functions of barrier height. A linear fit gives a slope of $0.0435\pi\ \mathrm{meV}^{-1}$, with $\pi$ rotation at $V_\pi \approx 23$ meV. The simulations use $E_{\mathrm{F}} = 80$ meV, $v_{\mathrm{F}} = 10^6\ \mathrm{m\ s^{-1}}$, and $\zeta_y = 0.35$.

## C. Time-resolved phase rotation

Because the barrier-induced relative phase must be dynamically extracted from the scattered wave packet, the choice of comparison time is critical. As shown in Fig. 3a, the relative phase evolves through four distinct regimes whose boundaries correspond to well-defined physical events. Region A is characterized by free propagation before the wave packet reaches the first barrier interface, during which time $\Delta\phi = 0$ as both valleys evolve identically. Region B spans the active barrier interaction, beginning when the leading edge of the wave packet contacts the entrance interface and ending when both valley wave packets have fully cleared the oblique exit; $\Delta\phi$ builds rapidly throughout this interval, with the $\pi$ operating point showing a transient negative excursion as the wave packet enters the barrier. Region C is the postscattering plateau, where all three phases have converged to stable, barrier-height-dependent values. Region D captures the late-time regime in which drain absorption depletes the wave packet. Within the plateau, the phase is highly stable: once the transmitted packets have cleared the exit interface, $\Delta\phi$ is stationary to a fraction of a milliradian at every operating point, as expected from the vanishing of the barrier matrix element in the plateau (see Appendix 5 and Supplementary Note 2).

The overlap magnitudes (shown in Fig. 3b) undergo dramatic transient dips during the barrier interaction, reaching values below 0.1 for the $\pi$ operating point before recovering to stable plateau values. This transient suppression is expected: while the wave packet straddles the barrier, the with-barrier and zero-barrier spinors occupy largely nonoverlapping spatial regions, producing a momentary near-orthogonality that does not bear on the postscattering phase operation. Consequently, the optimal phase-extraction window is the subpicosecond plateau interval within region C. The same stability holds for the directly calculated cross-valley phase, whose temporal resolution reveals that the relative phase forms entirely within the barrier-traversal window (see Supplementary Note 3).

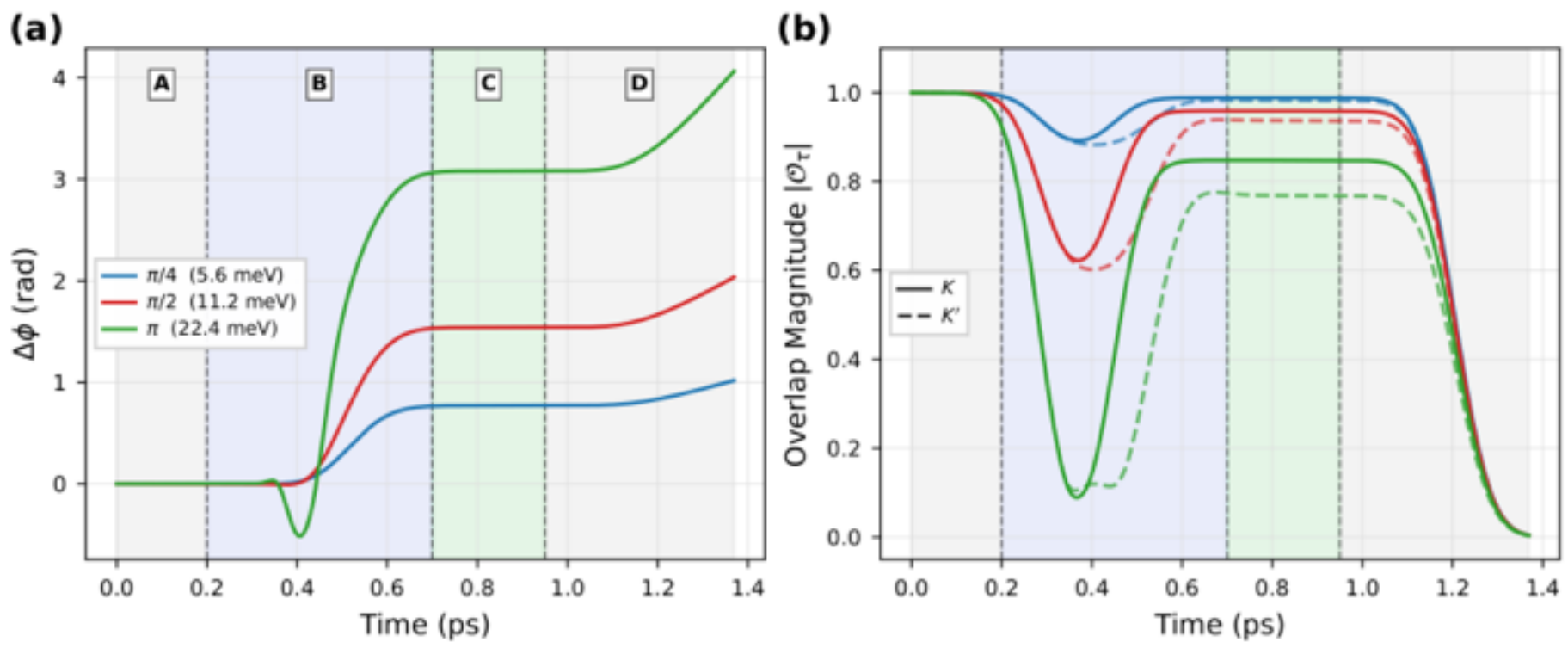

FIG. 3. **The time-resolved phase and overlaps explain the reference-based extraction window.** **a**, Time dependence of the relative valley phase $\Delta\phi(t)$ for the three representative operating points: $\pi/4$ at 5.6 meV (blue), $\pi/2$ at 11.2 meV (red), and $\pi$ at 22.4 meV (green). Shaded regions delineate four temporal regimes labeled A–D: free propagation, active barrier interaction, the postscattering plateau, and late-time drain absorption. **b**, Time dependence of the per-valley overlap magnitude $|\mathcal{O}_\tau(t)|$ for $K$ (solid) and $K'$ (dashed). During the barrier interaction (region B), the overlap undergoes a deep transient dip—most pronounced for the $\pi$ operating point, where $|\mathcal{O}_K|$ decreases to 0.09 and $|\mathcal{O}_{K'}|$ to 0.11—as the with-barrier spinor is momentarily near-orthogonal to the zero-barrier reference while the wave packet straddles the potential. After the scattering is complete, the overlap recovers and saturates within the plateau at $|\mathcal{O}_K| = 0.99,\ 0.96,\ 0.85$ and $|\mathcal{O}_{K'}| = 0.98,\ 0.94,\ 0.77$ for the $\pi/4$, $\pi/2$, and $\pi$ operating points, respectively. The phase used in the barrier-height dependence (Fig. 2) is extracted from the plateau (region C), where the wave packet has cleared the oblique exit interface but has not yet reached the drain absorber.

## D. Coherent valley-dependent phase accumulation

The real-space signatures consistent with the overlap-extracted phase are illustrated in Fig. 4, which plots the normalized real-part line-outs of the positive-energy projected wave function, $\mathrm{Re}[\tilde{\psi}]$, along the dominant transmitted transverse coordinate for each valley. Each panel overlays the $K$ and $K'$ line-outs, extracted at opposite transverse positions set by the bare tilt-induced drift already present in the zero-barrier reference. At the $\pi/4$ operating point, the two line-outs exhibit nearly identical spatial envelopes with a phase displacement consistent with the $\pi/4$ target. Raising the barrier to the $\pi/2$ and $\pi$ operating points yields line-outs consistent with those targets.

The line-outs evolve from a barely perceptible phase offset to nearly antiphased carrier oscillations. The retained-mode probabilities (see Fig. 3b) show that the $\pi/4$ and $\pi/2$ cases remain close to the transported reference modes, whereas the $\pi$ point, at the upper boundary of the tunability range, already shows appreciable loss of reference-mode weight. The device thus accumulates an order-$\pi$ barrier-induced relative valley phase before substantial deviation from the transported reference mode sets in.

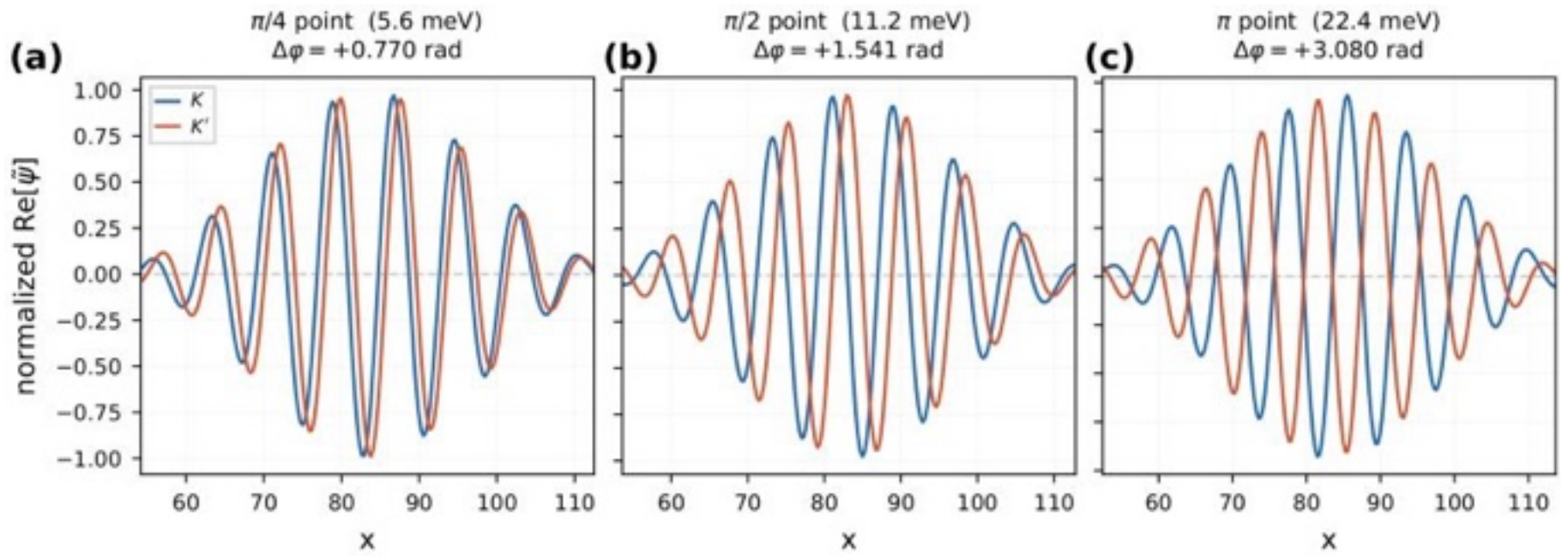

FIG. 4. **Waveform-level signatures of transported valley phase accumulation.** Normalized real-part line-outs of the positive-energy projected wave function, Re$[\tilde{\psi}]$, along the dominant transmitted coordinate for three barrier heights. **a**, $\pi/4$ point: the valleys have nearly identical envelopes with a phase shift at the $\pi/4$ target. **b**, $\pi/2$ point: the phase shift creates a quarter-cycle offset with preserved envelopes. **c**, $\pi$ point: the phase produces nearly antiphased oscillations. The envelope shows minor amplitude modulation, indicating the onset of coherent deviation. The extracted phase shifts and barrier heights are annotated in each panel; the residuals from the exact targets are set by the barrier-height grid step. All wave function line-outs are shown at a fixed instant within the postscattering plateau (region C of Fig. 3).

## E. Ideal operating regime

The breakdown of the ideal low-barrier operating regime as the barrier height increases can be traced to a coherent geometrical divergence of the wave packet rather than to phase randomization. This distinction is made clear by examining the four independent diagnostics presented in Fig. 5. The first effect is observed in the principal-axis angle of the transmitted wave-packet density (see Fig. 5a), which shows that the barrier gradually transforms the wave packet from its initial transversely elongated Gaussian profile (with $\theta = 90°$) into a longitudinally stretched envelope. This reshaping becomes strongly valley dependent; for example, midway through the reflection crossover, the principal-axis orientations for the two valleys differ by $\approx 20°$. This angular separation provides a direct real-space manifestation of the coherent reshaping process, which, in turn, reduces the overlap between the barrier-transmitted and reference wave packets, as defined in Eq. (4). The second effect is captured by the normalized current circulation (shown in Fig. 5b), which vanishes at zero barrier height but develops oppositely signed vortical components for the two valleys as the barrier induces asymmetric current flow—again coherent, valley-dependent deformation rather than incoherent scattering.

The third diagnostic, the valley polarization (shown in Fig. 5c), remains negligible up to and including the $\pi$ point; a finite polarization ($\eta \approx +0.05$) emerges only as the barrier height approaches $E_{\mathrm{F}}$, where transmission becomes both suppressed and valley dependent. The fourth diagnostic, centroid position analysis (presented in Fig. 5d), provides additional insight into the degradation mechanism. Here, the transverse centroid shift $y_c - y_c(0)$ increases monotonically with $V_0$ for both valleys, which is a consequence of oblique-exit refraction that steers both wave packets in the same transverse direction. The longitudinal centroid shift is negligible in the low-barrier regime but decreases sharply near $E_{\mathrm{F}}$, where the barrier significantly delays and restructures the transmitted wave packet. As $V_0$ approaches $E_{\mathrm{F}}$, all four metrics deviate sharply from ideal phase-element behavior, marking the transition to a regime in which both phase and filtering effects are significant.

The tradeoff between phase tunability and reference-mode leakage is made explicit in Fig. 6c (presented alongside the interferometric readout in the Discussion): a 5% mean-leakage budget admits rotations up to $\approx \pi/3$ and a 10% budget up to $\approx \pi/2$, whereas a $\pi$ rotation for this geometry costs roughly a third of the reference-mode weight. Near the origin, the tradeoff is parabolic, $L \propto (\Delta\phi)^2$, as follows from the deflection-dominated leakage law of Supplementary Note 2. Larger $\Delta D_{\mathrm{eff}}$—higher tilt or optimized interface angle, Eq. (D1)—lowers $V_\pi$ and therefore the leakage cost of a given rotation.

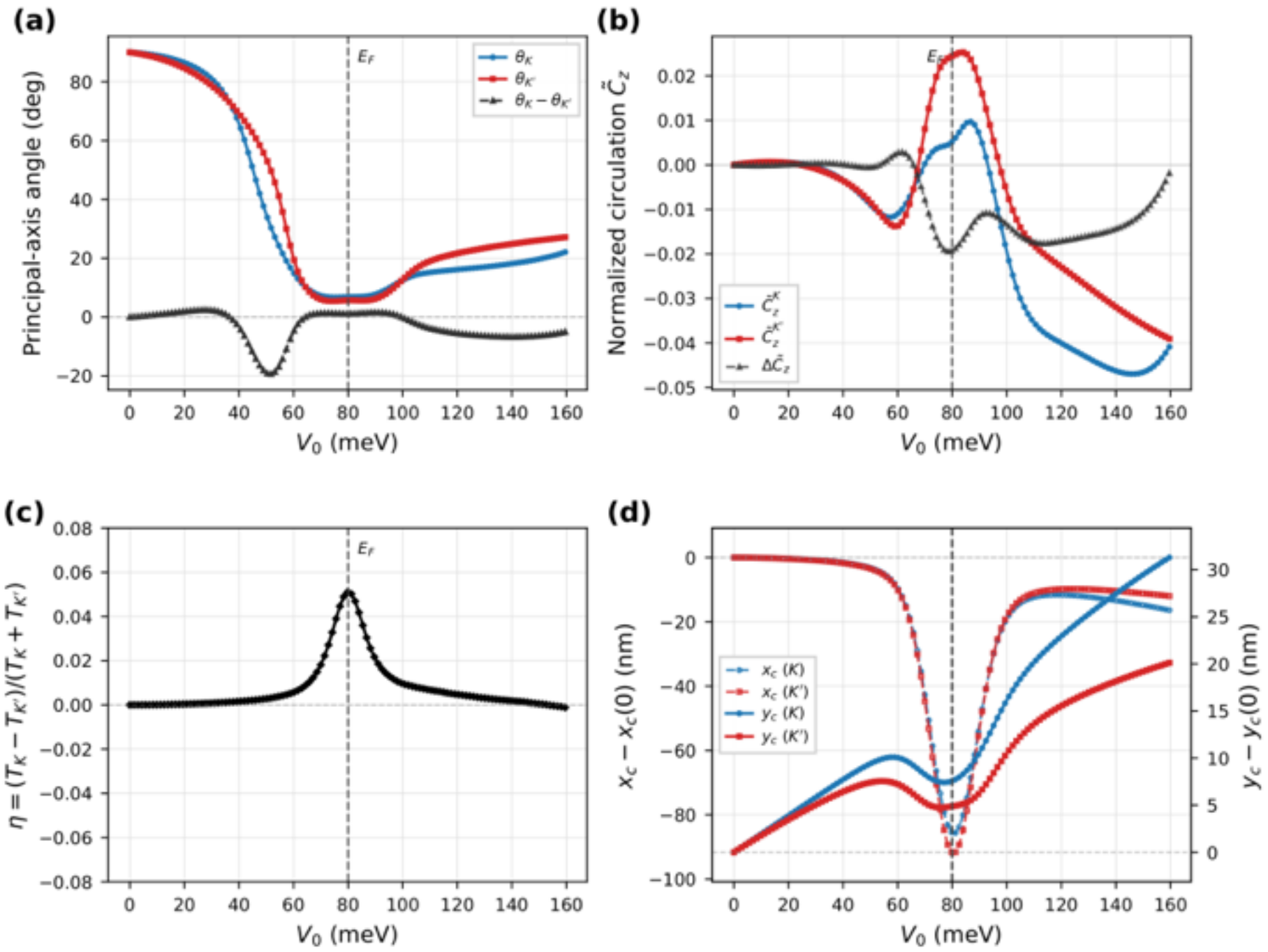

FIG. 5. **Geometric and transport diagnostics of coherent deviation from the transported reference modes.** **a**, Principal-axis angle of transmitted wave packet density for $K$ (blue) and $K'$ (red) valleys versus barrier height. Both start at 90°; as $V_0$ increases, the axes rotate toward propagation, reaching minima near $E_{\mathrm{F}}$. The valley angle difference remains $\lesssim 2°$ across the low-barrier operating window but decreases to $-19°$ near $V_0 = 52$ meV. **b**, Normalized current circulation $\tilde{C}_z$ for each valley and their difference versus $V_0$. Circulation is negligible at low barrier heights, peaking near $E_{\mathrm{F}}$ because of the large Fermi-surface mismatch between oppositely tilted cones in different regions. **c**, Valley polarization $\eta$ of the transmitted current. $\eta$ is nearly zero for low barriers ($|\eta| < 5 \times 10^{-4}$ up to the $\pi$ point), but reaches $0.05$ near $E_{\mathrm{F}}$. **d**, Centroid position of the transmitted wave packet versus the barrier height relative to the initial position. Both valleys shift transversely by different amounts, with separation $\Delta y_c$ increasing with $V_0$. Near $E_{\mathrm{F}}$, the longitudinal centroid decreases sharply due to delayed, reshaped transmission. In all panels, the dashed vertical line indicates $E_{\mathrm{F}} = 80\,\mathrm{meV}$.

## III. DISCUSSION

The central result is the demonstration of a coherent, electrically tunable valley-phase operation in a tilted Dirac or Weyl channel. The geometric tilt of the cone generates a continuously tunable relative dynamical phase without the need for engineered valley energy splitting. The tilt serves two roles: it separates the valleys transversely during propagation, establishing the transported zero-barrier references as the natural comparison basis, and, within that basis, it enables the shaped electrostatic barrier to function as a controllable phase element. The accumulated phase, while fundamentally dynamical and not simply a ray-optic path length, can, in the low-barrier regime, be interpreted using an effective-width framework determined by the device geometry.

### A. Operational definition and interferometric readout

The physical content of the operation is fixed by three computed quantities: the transmissions $T_K$ and $T_{K'}$, the transported-mode overlaps $\mathcal{O}_\tau$, and the intervalley coherence $\mu$. The last is directly accessible in an interferometric geometry (shown in Fig. 6a): valley-polarized injection,[12,14,18] a valley-mixing stage preparing $(|K\rangle + |K'\rangle)/\sqrt{2}$ (candidate implementations include line-defect scatterers[14] and engineered short-range mixing regions), the phase element studied here, a recombining mixer, and valley-resolved detection.[19] Because the charge current has no intervalley matrix elements, the output populations are determined by the triple $(T_K, T_{K'}, \mu)$ alone, and the predicted conductance oscillates as $\cos^2(\Delta\phi(V_0)/2)$ with period $2\pi/(\mathrm{d}\Delta\phi/\mathrm{d}V_0) = 2\pi\hbar v_\mathrm{F}/\Delta D_\mathrm{eff} \approx 0.57\, E_\mathrm{F}$ for this geometry—a specific experimental signature (shown in Fig. 6b). The computed visibility envelope follows the deflection-model law $1 - |\mu|^2 = L_0 + cV_0^2$ (see Supplementary Note 3). Supplementary Note 4 presents an independent dc signature: the transmitted valley current-angle splitting grows in the same $1:2:4$ ratio as the phase targets.

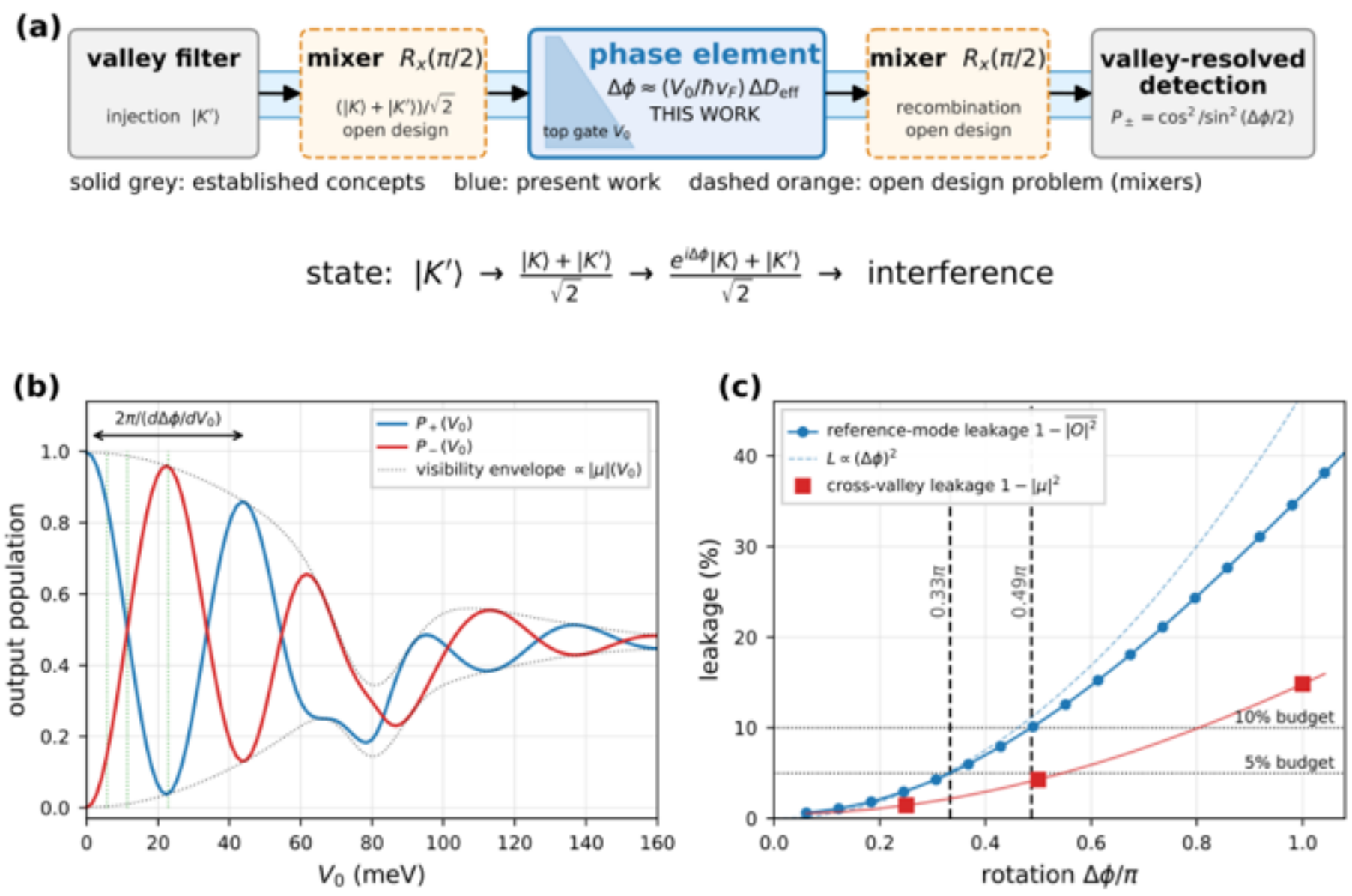

FIG. 6. **Interferometric readout of the valley phase.** **a**, Two-path valley interferometer embedding the phase element: valley-polarized injection, a mixing stage preparing $(|K\rangle + |K'\rangle)/\sqrt{2}$, the electrostatic phase element (present work), a recombining mixer, and valley-resolved detection. The solid gray blocks are established concepts; the dashed blocks (mixers) are open design problems. **b**, Readout constructed entirely from the triple $(T_K, T_{K'}, \mu)$: the output population oscillates as $\cos^2(\Delta\phi(V_0)/2)$ with the period annotated in the panel, bounded by the computed visibility envelope $|\mu|(V_0)$. **c**, Tunability–leakage tradeoff: reference-mode leakage $1 - \overline{|\mathcal{O}|^2}$ versus rotation angle, with the parabolic law $L \propto (\Delta\phi)^2$ and the cross-valley leakage $1 - |\mu|^2$ lying less than half of it.

## B. Material realization and experimental feasibility

While our model is formulated for a generic tilted Dirac/Weyl fermion platform, several materials possess the requisite parameters for experimental realization. Among two-dimensional systems, the 8-$Pmmn$ phase of borophene[20,21] is a prominent, theoretically predicted candidate hosting highly tilted, anisotropic Dirac cones.[22] Experimentally verified tilted Dirac cones in organic conductors and suitably chosen Weyl-semimetal flakes with an effective in-plane type-I tilt provide additional candidate platforms, as discussed in Supplementary Note 5. Given a typical Fermi velocity of $v_{\mathrm{F}} \sim 10^6$ m s$^{-1}$ and the few-hundred-nanometer device dimensions modeled here, the required barrier heights ($V_\pi \approx 0.3\,E_{\mathrm{F}}$, i.e., meV-scale) fall well within standard electrostatic gating capabilities. The mass-wall confinement used here is an idealized continuum boundary model; a material-specific implementation would require engineered lateral confinement or gapped edges, with valley-preserving termination (zigzag-type or a smooth induced gap) as an explicit design requirement.

### C. Robustness against disorder

In practical devices, static spatial disorder is unavoidable, with charge puddles in the top-gate oxide serving as a representative example. To assess the experimental feasibility of our proposal, we have incorporated static spatial potential fluctuations characterized by a root-mean-square amplitude of $5\%$ relative to $V_0$ and a finite correlation length, as described in Supplementary Note 6. The first effect of introducing such disorder is a steepening of the phase-versus-barrier-height slope by approximately $12\%$, which, in turn, shifts the required operating barrier heights downward by approximately $9\%$. The overlap magnitudes remain essentially unchanged from the disorder-free case. Across the entire low-barrier operating window, the transmission probabilities remain close to unity, and the valley polarization is maintained below $5 \times 10^{-4}$. Because such disorder is smooth on the lattice scale, it does not activate the intervalley coupling channel of Eq. (1). Any phase offset induced by smooth disorder can therefore be compensated by recalibrating the macroscopic top gate.

### D. Robustness against intervalley coupling

Beyond smooth disorder, the coupled two-valley model quantifies the short-range channel directly (see Supplementary Note 1). A scalar intervalley field of strength $u$ confined to the gated region produces single-pass transfer $p = \sin^2(u\, t_{\text{eff}}/\hbar)$, with $t_{\text{eff}} = 194\,\text{fs}$ ($K$) and $251\,\text{fs}$ ($K'$) set by each valley's traversal chord through the coupling region. The transfer is independent of barrier height—massless carriers traverse at $v_{\text{F}}$ regardless of $V_0$, so the exposure is geometric, not energetic—and directionally asymmetric, $p(K' \to K)/p(K \to K') = 1.65$, through the same valley-dependent dwell that generates the phase. Percent-level operation requires $u \lesssim 0.26\,\text{meV}$: up to this threshold, both the transported-mode overlap and the directly calculated cross-valley coherence at the $\pi$ point are unchanged within tolerances, so the barrier continues to act as a pure valley-phase element, with appreciable degradation setting in only near the $10\%$ threshold ($u \approx 0.8\,\text{meV}$). Expressed as an experimentally accessible criterion, the $1\%$ threshold corresponds to a ballistic intervalley length $L_{\text{iv}} = v_{\text{F}}\tau_{\text{iv}} \approx 20\ \mu\text{m}$; because the intervalley scattering time $\tau_{\mathrm{iv}}$ is routinely determined by weak-localization magnetotransport (see Supplementary Note 1 for the conversion between the ballistic and diffusive length conventions), this is a concrete sample-selection rule.

### E. Toward full valley-coherent control

While this work isolates the phase-accumulation stage, full coherent control of the valley degree of freedom would also require basis mixing and readout. Because the present barrier geometry is designed to suppress intervalley scattering, a separate coherent recombination or purpose-designed valley-mixing stage is needed to convert the transported-reference phase into a drain-local valley-only rotation and to realize complementary mixing operations. The coupled-valley simulations of Supplementary Note 1 indicate that such stages are physically available: engineered intervalley-mixing regions drive coherent Rabi-type valley rotations whose contrast is electrically frozen during copropagation. Combined with established concepts for valley-Hall initialization and readout,[19] the present electrostatic phase element could serve as the fast phase-control component of a broader valley-coherent architecture.

## SUPPLEMENTARY MATERIAL

See the supplementary material for eight supplementary notes (with eight figures and three tables) supporting the main text: robustness against coherent intervalley coupling (Note 1); the transported-reference phase extraction and diagnostics of orbital mismatch (Note 2); the directly calculated intervalley coherence and validation of the reference-based extraction (Note 3); the time-resolved probability current (Note 4); tilted energy dispersions and candidate material platforms (Note 5); robustness against static spatial disorder (Note 6); details of the split-operator Fourier propagation method (Note 7); and the analytical approximation of the phase model (Note 8).

## ACKNOWLEDGMENTS

This work was funded by the Nanoelectronics Research Center (Nano Electronics Research Mühendislik Araştırma Geliştirme Ltd. Şti.), Istanbul, Turkey.

## AUTHOR DECLARATIONS

### Conflict of Interest

The author has no conflicts to disclose.

### Ethics Approval

Ethics approval is not required.

### Author Contributions

C.Y. conceived the project, developed the numerical calculation and analysis methods, performed the calculations, analyzed the data, and wrote the manuscript.

## DATA AVAILABILITY

The data underlying the figures and analyses were generated by numerical calculation with the code described below; the exact datasets are available from the author upon reasonable request. The wave-packet simulations were performed with the open-source `dirac-wavepacket` package, version 1.0.1,[26] openly available on Zenodo at https://doi.org/10.5281/zenodo.19651771 with no restrictions on access.

### Appendix A: Model and propagation method

All results are obtained from coherent time-dependent calculations on the coupled two-valley Dirac/Weyl model of Eq. (1), with reflecting sidewalls and absorbing drains. The barrier geometry is specified by two linear interfaces [Eq. (8)], yielding a vertical entrance and an oblique exit. The injected state in each valley block is a Gaussian positive-energy spinor packet,

$$\Psi^{\text{in}}(x, y) = \mathcal{N}\exp\left[-\frac{(x - x_0)^2}{4\sigma_x^2} - \frac{(y - y_0)^2}{4\sigma_y^2}\right] e^{i(k_{x0}x + k_{y0}y)} \chi_+(k_{x0}, k_{y0}), \tag{A1}$$

where $\chi_+ = 2^{-1/2}(1, e^{i\tau\phi_0})^{\mathrm{T}}$, $\phi_0 = \arg(k_{x0} + ik_{y0})$, $\tau = \pm 1$ is the valley index, and $\mathcal{N}$ is the normalization constant of the Gaussian spinor packet. Time-dependent Dirac evolution is carried out using a second-order split-operator Fourier method,[23–25] which alternately applies real-space and momentum-space propagators to evolve the four-component spinor. Physical units are obtained by mapping $E_{\mathrm{F}} = \hbar v_{\mathrm{F}} k_0 = 80$ meV and $v_{\mathrm{F}} = 10^6$ m s$^{-1}$. Splitting the Hamiltonian of Eqs. (1) and (2) as $H = H_r + H_k$, where $H_r$ collects the position-dependent terms [the barrier $V(x, y)$, the confinement $M(y)\sigma_z$, and the intervalley coupling $U_{KK'}$] and $H_k$ the valley-diagonal kinetic and tilt terms, one time step is approximated as follows:

$$\Psi(t + \Delta t) \approx e^{-iH_r\Delta t/2\hbar} \mathcal{F}^{-1} e^{-iH_k\Delta t/\hbar} \mathcal{F}\, e^{-iH_r\Delta t/2\hbar} \Psi(t), \tag{A2}$$

with local error $O(\Delta t^3)$, where $\mathcal{F}$ and $\mathcal{F}^{-1}$ denote the forward and inverse two-dimensional Fourier transforms, respectively. The momentum-space step factorizes into the two valley blocks and is evaluated analytically as a $2 \times 2$ unitary for each valley at every $(k_x, k_y)$ point, with the valley's own tilt sign and chirality. The real-space half step, in which $U_{KK'}$ acts, is local and factorizes into exact $2 \times 2$ mixing exponentials for each pseudospin component; at $U_{KK'} = 0$, the scheme reduces identically to two independent single-valley propagations. For the waveform plots in Fig. 4, we display the scalar amplitude obtained by projecting the propagated spinor onto the injected positive-energy eigenspinor, $\tilde{\psi}(x, y, t) = \chi_+^{\dagger}(k_{x0}, k_{y0})\Psi(x, y, t)$. The details of the calculation and boundary-condition specifications are provided in Supplementary Note 7.

## Appendix B: Transport observables and phase extraction

Transmission and reflection probabilities are computed from the cumulative probability absorbed by the drain masks or, equivalently, from the current flowing into the absorbing drains. For valley $\tau$, the transmission $T_\tau$ and reflection $R_\tau$ satisfy $T_\tau + R_\tau + P_\tau^{\mathrm{rem}} = 1$, where $P_\tau^{\mathrm{rem}}$ is the remaining probability in the channel. To extract the barrier-induced phase, we compare the barrier-propagated spinor at a chosen comparison time $t^*$ with a reference spinor evolved without the barrier under otherwise identical conditions, yielding the overlap $\mathcal{O}_\tau$ of Eq. (4), whose magnitude serves as a mode-preservation diagnostic; the relative valley phase $\Delta\phi$ follows from Eq. (5). The comparison time $t^*$ is chosen within the postscattering plateau where the transmitted packets have cleared the oblique interface but have not yet been substantially absorbed, as identified in Fig. 3; within the plateau, the extraction is $t^*$-independent by construction, since once the packets carry negligible weight on the barrier, $(\mathrm{d}/\mathrm{d}t)\langle\Psi^{(0)}|\Psi^{(V_0)}\rangle \propto \langle\Psi^{(0)}|V_{\mathrm{barrier}}|\Psi^{(V_0)}\rangle \approx 0$ (see Supplementary Note 2). Additional diagnostic definitions, including the principal-axis angle, centroid shift, and circulation metrics of Fig. 5, are detailed in the supplementary material.

### Appendix C: Cross-valley coherence

The intervalley coherence $\mu = \langle\chi_{K'}|\chi_K\rangle$ is computed between the transmitted (downstream of the exit interface) and individually normalized valley states at the plateau step after the $K'$ centroid is aligned onto $K$ by a subpixel Fourier shift; the alignment represents the recombination that a mixer performs and is the only convention entering $\mu$. At $V_0 = 0$, this yields $|\mu|$ equal to unity within a fraction of a percent; the residue, set by the spinor convention (see Supplementary Note 3), fixes the small baseline $L_0$ of the visibility envelope.

### Appendix D: Analytical approximation

An approximate analytical estimate of the relative phase is obtained by evaluating the longitudinal propagation constant of a valley-$\tau$ Dirac carrier inside a tilted barrier. For a uniform rectangular barrier, the longitudinal wavevector is $q_{x,\tau} = \frac{1}{\hbar v_{\mathrm{F}}}\sqrt{(E - V_0 + \tau\hbar w_y k_y)^2 - (\hbar v_{\mathrm{F}} k_y)^2}$ such that each valley accumulates a dynamical phase $\phi_\tau = q_{x,\tau} D$ over a barrier length $D$. For shaped interfaces, the accumulated phase becomes a line integral, $\phi_\tau = \int q_{x,\tau}(y)\,\mathrm{d}x$. The difference $\Delta\phi = \phi_K - \phi_{K'}$ thus reflects the difference in dwell time between valleys and reduces to Eq. (9) in the low-barrier limit. The effective widths entering Eq. (9) follow in closed form from the group-velocity ray $y_\tau(x) = \tau(w_y/v_{\mathrm{F}})(x + d)$ intersected with the oblique exit interface:

$$D_\tau = \frac{W\,(L_y/2 - \tau\zeta_y d)}{L_y + \tau\zeta_y W}, \qquad \Delta D_{\rm eff} = \frac{\zeta_y \tan\alpha\,(W + 2d)}{1 - \zeta_y^2 \tan^2\alpha}, \qquad \tan\alpha = \frac{W}{L_y}, \qquad \text{(D1)}$$

where $\zeta_y = w_y/v_{\rm F}$ is the tilt parameter, $W$ is the barrier width, $d$ is the propagation distance from the injection to the entrance interface, $L_y$ is the channel width, and $\alpha$ is the exit-interface angle. Equivalently, $\Delta\phi = V_0\,\Delta t_{\rm dwell}/\hbar$ with $\Delta t_{\rm dwell} = \Delta D_{\rm eff}/v_{\rm F}$: the phase is the electrostatic energy accumulated over the valley dwell-time asymmetry. For the device geometry, $\Delta D_{\rm eff} \approx$ 90 nm, reproducing the numerical low-barrier slope shown in Fig. 2d to within a percent. The closed form makes the design scalings explicit: $\Delta\phi$ is linear in $\zeta_y$ for $\zeta_y \tan\alpha \ll 1$, with geometric enhancement $(1 - \zeta_y^2 \tan^2\alpha)^{-1}$, and linear in the interface angle at small $\alpha$; simulations spanning four tilts ($\zeta_y = 0.15$–$0.45$) and four interface angles ($\alpha = 8.5°$–$16.7°$) reproduce Eq. (D1) to within $+0.5$–$0.7\%$ across a factor-8 range of slopes (see Supplementary Note 8). The analytical curves in Fig. 2 are used to evaluate the full phase integral without invoking classical ray trajectories; the complete derivation is provided in Supplementary Note 8.

# Supplementary Material

Electrostatic control of valley-dependent phase in tilted Dirac/Weyl channels

Can Yesilyurt
Nanoelectronics Research Center, Istanbul, Turkey

July 28, 2026

This supplementary material comprises eight supplementary notes, eight supplementary figures, and three supplementary tables.

## Contents

# 1 Supplementary Note 1: Robustness against coherent intervalley coupling

This section quantifies the effect of the intervalley coupling field $U_{KK'}$ of Eq. (17) on the phase operation. A scalar coupling $U_{KK'} = u\,\mathbb{I}$ is considered, confined to the gated region and matching the barrier footprint. The coupled system is simulated for coupling strengths spanning $u = 0.001$–$5\,\mathrm{meV}$, for both injection valleys and across the operating window of barrier heights. Supplementary Fig. 1 summarizes the results.

***Transfer law and thresholds.*** The single-pass intervalley transfer follows a coherent Rabi law over more than four decades,

$$p = \sin^2\left(\frac{u\,t_{\mathrm{eff}}}{\hbar}\right), \qquad t_{\mathrm{eff}} = 194\,\mathrm{fs}\ (K\ \text{injection}), \quad 251\,\mathrm{fs}\ (K'\ \text{injection}), \tag{1}$$

The effective interaction times are determined by each valley's traversal chord through the coupling region and are approximately in the ratio of the geometric chords described in Supplementary Note 8. The transfer remains independent of barrier height to within $< 0.2\%$: massless carriers traverse the gated region at $v_{\mathrm{F}}$ regardless of $V_0$, so the exposure to the coupling is geometric rather than energetic. The thresholds for percent-level and ten-percent-level transfer are

$$u_c(1\%) = 0.26\,\mathrm{meV}, \qquad u_c(10\%) \approx 0.8\,\mathrm{meV} \tag{2}$$

of coherent coupling sustained across the entire gated region. Beyond threshold, the operation degrades in the manner expected of coherent dynamics—a Rabi turnover is observed at $u = 5\,\mathrm{meV}$ for both injection directions (see Supplementary Fig. 1c,d)—rather than by stochastic quenching.

***Directional asymmetry and walk-off.*** The transfer exhibits directional asymmetry, $p(K' \to K)/p(K \to K') = 1.65$, which remains constant over four decades of coupling. Time-resolved conversion for the two injection directions is compared in Supplementary Fig. 1c,d: the $K$-injected packet exits the coupling region earlier and its conversion window closes sooner, consistent with the shorter traversal chord ($t_{\mathrm{eff}} = 194$ versus $251$ fs). This asymmetry reflects the same valley-dependent dwell responsible for generating the phase; the robustness limit and the operating principle are manifestations of a single mechanism. A converted amplitude inherits the target valley's tilt, resulting in transverse walk-off of the two components after conversion; coherent accumulation is truncated at the walk-off length $\ell_{\mathrm{wo}} \approx \sigma_y v_{\mathrm{F}}/(2w_y)$. This property ensures that interference contrast established at a mixing stage is preserved during subsequent copropagation, rather than being degraded by valley-dependent drift. This observation is critical for the recombination stage of the interferometric readout (see Fig. 6 of the main text).

***Sensitivity of the phase operation.*** The key operational question is whether the valley phase shift persists under intervalley coupling in the barrier region, including short-range scattering at the barrier junctions encompassed by the coupling field. Supplementary Fig. 1b addresses this directly at the $\pi$ operating point: the mean transported-mode overlap $|\mathcal{O}|$ (reference-based) and the cross-valley coherence $|\mu|$ (directly calculated, without reference evolution) remain unchanged within tolerances up to $u_c(1\%)$, degrade appreciably only beyond $u_c(10\%)$, and collapse by $u = 2\,\mathrm{meV}$, where the single-pass transfer reaches tens of percent. As both metrics—one extraction-based, one extraction-free—respond identically across the scan, this robustness is intrinsic to the phase operation itself rather than the extraction method. Below the percent threshold, the barrier continues to function as a pure valley-phase element with clean-device visibility; above this threshold, degradation follows the coherent Rabi dynamics of Eq. (1) rather than stochastic dephasing.

***Experimental criterion.*** The relevant materials parameter is directly accessible experimentally: the intervalley length $L_{\mathrm{iv}}$ obtained from weak-localization magnetotransport. Here, $L_{\mathrm{iv}} = v_{\mathrm{F}}\,\tau_{\mathrm{iv}}$ denotes the ballistic intervalley length appropriate to the present device; diffusive weak-localization analyses report the shorter length $\sqrt{D\,\tau_{\mathrm{iv}}}$ for the same intervalley time $\tau_{\mathrm{iv}}$, making the comparisons with measured values below conservative. In these terms, $p = 1\%$ over the ~200–250 fs dwell corresponds to $L_{\mathrm{iv}} \approx 19$–$25\ \mu$m, consistent with the analytic single-pass criterion ($L_{\mathrm{iv}} \gtrsim 30\ \mu$m for percent-level operation), and the 10% criterion corresponds to $L_{\mathrm{iv}} \approx 2\ \mu$m. Although micrometer-scale intervalley lengths are demanding, they are supported by the experimental record: weak-localization analyses of exfoliated graphene report intervalley lengths up to $\sim 1\ \mu$m that are largely *edge-limited* rather than bulk-limited [1], an extrinsic channel absent in the present geometry, where the packets remain more than $5\sigma$ from the confinement walls. Additionally, the observation of the chiral-anomaly-induced negative magnetoresistance in TaAs requires internode relaxation slower than intranode scattering [2]. The specific microscopic disorder that realizes a given $u$ is material-dependent; the engineered device elements themselves cannot provide the required $|K - K'|$ momentum transfer (see Supplementary Note 7), and interfacial intervalley transmission is kinematically excluded by the Fermi-surface no-overlap criterion, which is satisfied with several-fold margins by reported TaAs node separations (0.7–0.8 nm$^{-1}$) and by more than a factor of six in $Ta_3S_2$ (1.5 nm$^{-1}$, the largest among known Weyl candidates; a type-II material, cited here only for its record node separation) [3, 4].

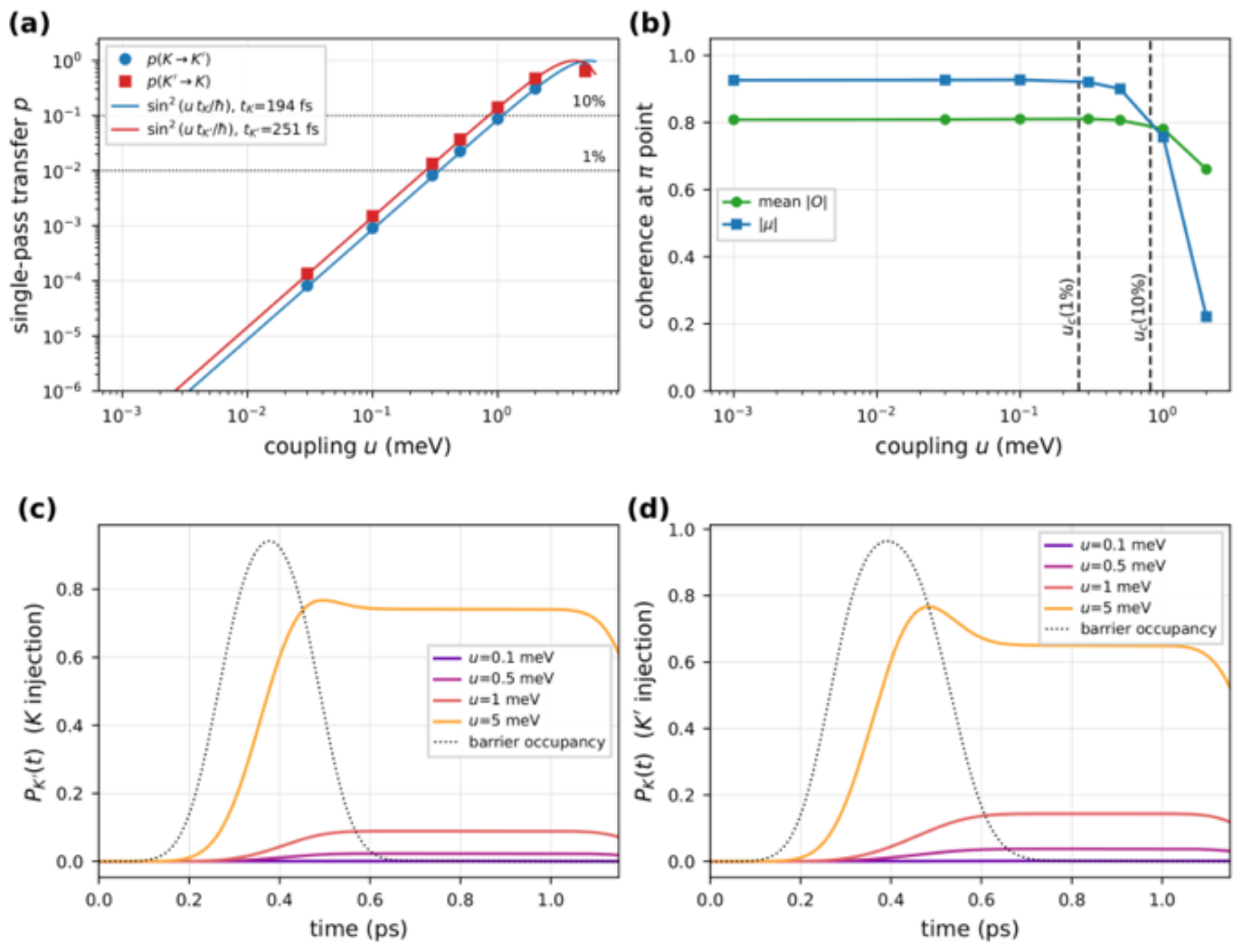

**Supplementary Fig. 1. Robustness of the phase operation against coherent intervalley coupling.** **a**, Calculated single-pass transfer $p$ versus coupling strength $u$ for both injection directions, with the Rabi law of Eq. (1). **b**, Operation quality at the $\pi$ point versus coupling strength: mean transported-mode overlap $|\mathcal{O}|$ and cross-valley coherence $|\mu|$, with the thresholds $u_c(1\%)$ and $u_c(10\%)$ marked. **c**,**d**, Time-resolved transfer for $K$ injection (**c**) and $K'$ injection (**d**): conversion is confined to the barrier-traversal window (dotted: barrier occupancy), with Rabi turnover at strong coupling.

## 2 Supplementary Note 2: Relative phase in the transported reference basis and diagnostics of orbital mismatch

This section provides the full definitions and derivations for the reference-based phase extraction, the overlap magnitudes, and the geometric diagnostics used to identify the onset of orbital mismatch at elevated barrier heights.

***Reference-based phase extraction.*** All reported numerical relative phases are extracted from barrier–reference overlaps, not from $k$-space transmission phase diagnostics. At the common post-interface step $t^*$ (selected within the plateau region C of Fig. 3 of the main text), the propagated wave function for a barrier of height $V_0$ is compared with a reference run with the same geometry and $V_0 = 0$. For each valley, the overlap integral is

$$\mathcal{O}_\tau(V_0;\, t^*) = \iint \mathrm{d}^2\mathbf{r}\; [\Psi_\tau^{(0)}(\mathbf{r},\, t^*)]^\dagger\, \Psi_\tau^{(V_0)}(\mathbf{r},\, t^*), \tag{3}$$

where $\Psi_\tau^{(0)}$ is the zero-barrier reference and $\Psi_\tau^{(V_0)}$ is the barrier-propagated spinor. The per-valley overlap phase is

$$\phi_\tau(V_0) = \arg \mathcal{O}_\tau(V_0;\, t^*), \tag{4}$$

and the corresponding overlap magnitude is $|\mathcal{O}_\tau(V_0;\, t^*)|$. The relative phase reported in the barrier-height-dependence figures (i.e., Fig. 2 of the main text) is the relative overlap phase:

$$\boxed{\Delta\phi(V_0) = \arg[\mathcal{O}_K(V_0;\, t^*)\; \mathcal{O}_{K'}^*(V_0;\, t^*)]}. \tag{5}$$

This definition cancels the common propagation phase shared by the barrier and reference runs, thereby more directly isolating the barrier-induced relative phase than a raw transmission-phase comparison.

***Relative operation in the transported reference basis.*** Because bare tilted propagation already produces opposite transverse drift for the two valleys, the barrier is most naturally characterized relative to the corresponding zero-barrier evolution. At the common postscattering time $t^*$, the barrier-propagated state for valley $\tau$ can be decomposed as

$$|\Psi_\tau^{(V_0)}(t^*)\rangle = \mathcal{O}_\tau(V_0; t^*)\, |\Psi_\tau^{(0)}(t^*)\rangle + |\delta\Psi_\tau^{\perp}(t^*)\rangle, \qquad \langle\Psi_\tau^{(0)}(t^*) \mid \delta\Psi_\tau^{\perp}(t^*)\rangle = 0. \tag{6}$$

For normalized comparison wave functions, $\| \delta\Psi_\tau^{\perp} \| = \sqrt{1 - |\mathcal{O}_\tau|^2}$. The overlap $\mathcal{O}_\tau$, therefore, quantifies how much of the full propagated wave function remains in the transported reference mode for that valley. Within the two-dimensional transmitted reference subspace, the barrier-induced relative operation is approximately

$$U_{\mathrm{rel}}^{\mathrm{eff}}(t^*) \approx \begin{pmatrix} \mathcal{O}_K & 0 \\ 0 & \mathcal{O}_{K'} \end{pmatrix}. \tag{7}$$

When $|\mathcal{O}_K|$ and $|\mathcal{O}_{K'}|$ are both close to unity, the barrier reduces, up to a global phase, to a nearly pure valley-phase shift element with relative phase $\Delta\phi$. Because the overlap is evaluated on the full propagated wave function, its reduction has distinct origins in different barrier-height windows. For the present parameters, $|\mathcal{O}_\tau|$ remains chiefly a transmitted-reference-mode diagnostic for $V_0 \lesssim 25\,\mathrm{meV}$. Between $25\,\mathrm{meV}$ and $E_{\mathrm{F}}$, growing reflection lowers the full-wave-function overlap. For $E_{\mathrm{F}} < V_0 < 2E_{\mathrm{F}}$, Klein-regime negative refraction at the entrance and exit interfaces bends the transmitted path away from the zero-barrier reference at both interfaces, effectively doubling the transverse deflection relative to the reference and further reducing $|\mathcal{O}_\tau|$.

***Rationale for excluding intervalley orbital overlap as the primary metric.*** The absolute intervalley orbital overlap at the comparison plane addresses whether the two transmitted valleys already share a common orbital mode, allowing the state to be treated as a drain-local valley-only two-level system after tracing out orbital motion. In the current tilted geometry, this overlap can be significantly less than unity even at $V_0 = 0$ because the opposite bare tilt causes transverse separation of the valleys during free propagation. Therefore, this metric is not adopted as the primary diagnostic. Relying on it as the main metric would incorrectly fail the identity limit. The primary consideration for the present device is whether each barrier-propagated valley remains close to its respective transported zero-barrier reference mode, as quantified by $|\mathcal{O}_K|$ and $|\mathcal{O}_{K'}|$. Restoration of a common orbital mode would require a subsequent coherent recombination or valley-mixing stage.

***Selection of the comparison time.*** The comparison time $t^*$ is selected within the postscattering plateau (i.e., region C of Fig. 3 in the main text), where the transmitted wave packets have traversed the oblique exit interface but have not yet undergone significant absorption by the drains.

The same common post-interface step is used for the barrier and reference wave functions at every $V_0$, keeping the numerical phase on a common time reference across the computation.

Within this plateau, the phase remains highly stable: once the transmitted packets have cleared the exit interface, the extracted $\Delta\phi$ is stationary to a fraction of a milliradian across the plateau at every operating point (see Fig. 3a of the main text).

This stability is not merely numerical: within the plateau the extraction is $t^*$-independent by construction. Once the packets carry negligible weight on the barrier support, the barrier-propagated and reference states evolve under the same Hamiltonian on their common support, so

$$\frac{\mathrm{d}}{\mathrm{d}t}\langle \Psi_\tau^{(0)} \mid \Psi_\tau^{(V_0)}\rangle = \frac{i}{\hbar}\,\langle \Psi_\tau^{(0)} |\, V_{\mathrm{barrier}}\, | \Psi_\tau^{(V_0)}\rangle \approx 0, \tag{8}$$

and the overlap phase is stationary. This is why region C is a plateau at all: the phase accumulates in transit and is stored rather than generated at the analysis plane. The same stationarity holds for the directly calculated cross-valley phase $\arg\mu(t)$ (see Supplementary Note 3).

***Gauge and basis independence.*** The extraction is gauge-independent: under a global U(1) redefinition of the initial state, bra and ket acquire conjugate phases and $\mathcal{O}_\tau$ is unchanged, while a uniform potential offset $V \to V + c$ shifts $\phi_K$ and $\phi_{K'}$ equally and cancels in $\Delta\phi$; no vector potential is present. As numbers, $\mathcal{O}_\tau$, $\Delta\phi$, and $\mu$ are basis-independent inner products; what is basis-relative is the interpretation of the operation as acting on the valley pseudospin alone, which holds in the transported reference basis, while a drain-local valley-only rotation requires a downstream coherent recombination stage (see Fig. 6 of the main text).

***Decomposition of the reference-mode leakage.*** To identify the physical origin of the overlap reduction, the leakage $L = 1 - |\mathcal{O}_\tau|^2$ is separated through a ladder of overlaps: the full overlap $|\mathcal{O}|$; the transmitted-conditioned overlap $|\mathcal{O}_T|$, evaluated with a projector onto the region downstream of the oblique exit, which isolates the reflection share; the rigid phase-space-matched overlap $|\mathcal{O}_{\mathrm{disp}}|$, in which the barrier-propagated state is rigidly translated in position *and* momentum to align its first moments with the reference, which isolates the share carried by a rigid displacement of the transported mode; and the remainder, which quantifies genuine reshaping. Supplementary Table 1 lists the shares at the three operating points.

**Supplementary Table 1.** Decomposition of the reference-mode leakage $L = 1 - |\mathcal{O}_\tau|^2$ at the three operating points.

| Point | Valley | $L$ | Reflection share | Rigid-displacement share | Reshaping share |
|---|---|---|---|---|---|
| $\pi/4$ | $K$ / $K'$ | 0.020 / 0.032 | ≤ 0.0001 | 98.5% / 98.4% | 0.0002 / 0.0004 |
| $\pi/2$ | $K$ / $K'$ | 0.079 / 0.122 | ≤ 0.0005 | 98.5% / 98.4% | 0.0010 / 0.0016 |
| $\pi$ | $K$ / $K'$ | 0.287 / 0.419 | ≤ 0.0014 | 98.0% / 98.0% | 0.0046 / 0.0071 |

Three conclusions follow. First, reflection is negligible, consistent with $T > 0.99$ throughout the operating window. Second, $\sim 98\%$ of the reduction is a rigid displacement of the transported mode in phase space, dominated by the momentum (deflection) offset—the extracted $\delta k_y$ reaches $-0.052$ ($K$) and $-0.062$ ($K'$) at the $\pi$ point, matching the current-deflection angles of Supplementary Note 4—with genuine reshaping below 0.7% even at $\pi$. Notably, position-only alignment *lowers* the overlap:

the position and momentum offsets are correlated and must be matched jointly. Third, the ordering of the two valleys is explained: $K'$ loses overlap faster, although $K$ shows the larger centroid shift, because $K'$ carries the larger *momentum* offset and the overlap is deflection-dominated. The leakage is therefore coherent and geometric—which-path information carried by the propagation direction—not decoherence; and since $V_0$ is a classical control parameter, this orbital distinguishability between barrier settings does not dephase the valley state at fixed $V_0$.

***Transport observables.*** The final transmission probability for each valley is the total probability absorbed by the right drain:

$$T_\tau^{\mathrm{final}} = T_\tau(t \to \infty). \tag{9}$$

The valley polarization of the transmitted current is

$$\eta(V_0) = \frac{T_K^{\mathrm{final}} - T_{K'}^{\mathrm{final}}}{T_K^{\mathrm{final}} + T_{K'}^{\mathrm{final}}}. \tag{10}$$

In the low-barrier operating window, $|\eta| < 5 \times 10^{-4}$ at the $\pi$ operating point, consistent with the device acting primarily as a phase element rather than a valley filter.

The valley-dependent transmitted group delay is extracted from the drain-absorption histories:

$$\tau_{T,\tau} = \frac{\int \mathrm{d}t\ t\ \dot{T}_\tau(t)}{T_\tau^{\mathrm{final}}}, \qquad \tau_{T,\tau}^{\mathrm{peak}} = \arg\max_t \dot{T}_\tau(t), \qquad \Delta\tau_T = \tau_{T,K}^{\mathrm{peak}} - \tau_{T,K'}^{\mathrm{peak}}. \tag{11}$$

These quantities are not used to define the relative phase but provide a compact summary of the transport symmetry between the two valleys.

***Wave-packet geometry.*** To diagnose why the overlap magnitudes decrease at larger $V_0$, the transmitted wave-packet geometry is analyzed at the selected post-interface step. For the normalized density $\rho_\tau = \Psi_\tau^\dagger \Psi_\tau$, the wave-packet centroid is

$$\mathbf{r}_{c,\tau} = \frac{\iint \mathrm{d}^2\mathbf{r}\ \mathbf{r}\ \rho_\tau(\mathbf{r}, t^*)}{\iint \mathrm{d}^2\mathbf{r}\ \rho_\tau(\mathbf{r}, t^*)}, \tag{12}$$

and the covariance matrix is

$$\Sigma_{ij,\tau} = \frac{\iint \mathrm{d}^2\mathbf{r}\ (r_i - r_{c,i,\tau})(r_j - r_{c,j,\tau})\ \rho_\tau(\mathbf{r}, t^*)}{\iint \mathrm{d}^2\mathbf{r}\ \rho_\tau(\mathbf{r}, t^*)}. \tag{13}$$

From the eigenvalues of $\Sigma_\tau$, we extract the major and minor widths ($\sigma_{\mathrm{major}}, \sigma_{\mathrm{minor}}$), the aspect ratio, and the principal-axis angle $\theta_\tau$. These are the quantities plotted in Fig. 5a of the main text, which shows that the principal-axis angle rotates from 90° (transversely elongated, reflecting the injected Gaussian shape) toward $\sim 5°$–$7°$ (longitudinally stretched) near $V_0 = E_\mathrm{F}$, with a valley-angle difference reaching $-19°$ at $V_0 \approx 52$ meV.

***Current circulation.*** $C_{z,\tau}$ integrates the vortical content of the transmitted probability current:

$$C_{z,\tau} = \iint \mathrm{d}^2\mathbf{r}\ [(x - x_{c,\tau})\,j_{y,\tau} - (y - y_{c,\tau})\,j_{x,\tau}], \tag{14}$$

and its normalized version $\tilde{C}_{z,\tau}$ is obtained by dividing by the wave-packet probability at the analysis step. The circulation vanishes at zero barrier height and develops opposite-sign vortical components for the two valleys as the barrier induces asymmetric current flow (see Fig. 5b of the main text).

# 3 Supplementary Note 3: Direct intervalley coherence and validation of the reference-based extraction

This section defines the directly calculated intervalley coherence $\mu$, presents its time resolution and barrier-height dependence, and uses it to validate the reference-based phase extraction of Supplementary Note 2.

***Definition and convention.*** The coherence is evaluated between the transmitted, individually normalized valley states at the plateau step,

$$\mu = \langle \chi_{K'} \mid \chi_K \rangle, \tag{15}$$

after aligning the $K'$ centroid onto $K$ by a subpixel Fourier shift. The alignment represents the recombination that a mixing stage performs and is the only convention entering $\mu$; without it, the raw overlap collapses because the bare tilt separates the valleys spatially (see the dotted curve in Supplementary Fig. 2a)—a geometric separation, not decoherence, which is precisely why transported or aligned metrics are required in this geometry (see Supplementary Note 2). At $V_0 = 0$, the aligned coherence is $|\mu| = 0.9976$ under the time-reversal-conjugate spinor convention used for the reported results, and unity to within $2 \times 10^{-6}$ under the alternative convention; this $\leq 0.3\%$ convention dependence sets the small baseline $L_0$ of the visibility envelope below and is negligible against the barrier-induced decay.

***Time resolution: the phase forms in transit.*** Supplementary Fig. 2a,b resolves $\mu(t)$ through the scattering process. The magnitude $|\mu|(t)$ dips while the packets straddle the barrier and recovers to a stable plateau, mirroring the transient behavior of the reference overlaps (see Supplementary Note 2). The phase $\arg\mu(t)$ forms *entirely within the barrier-traversal window* and is flat to milliradians across the plateau: the relative phase is accumulated in transit, as the mechanism asserts, not at the analysis plane.

***Visibility envelope and validation.*** Across the barrier-height range (see Supplementary Fig. 2c), the visibility follows the deflection-model law

$$1 - |\mu|^2 = L_0 + c\, V_0^2, \qquad c = 2.54 \times 10^{-4}\ \text{meV}^{-2}, \tag{16}$$

confirming the zero-parameter prediction of the leakage decomposition ($c = 2.68 \times 10^{-4}$ meV$^{-2}$; see Supplementary Note 2) to within a few percent; in normalized form, $c\, E_F^2 \approx 1.6$. At the $\pi/4$, $\pi/2$, and $\pi$ operating points, $|\mu| = 0.9929$, $0.9791$, and $0.9258$. The corresponding cross-valley leakage $1 - |\mu|^2$ (1.4%, 4.1%, 14.3%) lies a factor 2.1–2.5 below the reference-mode leakage, because the common-mode part of the barrier-induced deflection cancels between the valleys. Finally, the directly calculated phase validates the reference-based construction without any reference evolution: $\arg\mu$ reproduces $\Delta\phi$ to $+2.2$, $+6.3$, and $+26$ mrad at the three operating points ($< 1\%$ of $\pi$), and to below 1 mrad in the low-barrier limit (see Supplementary Fig. 2d).

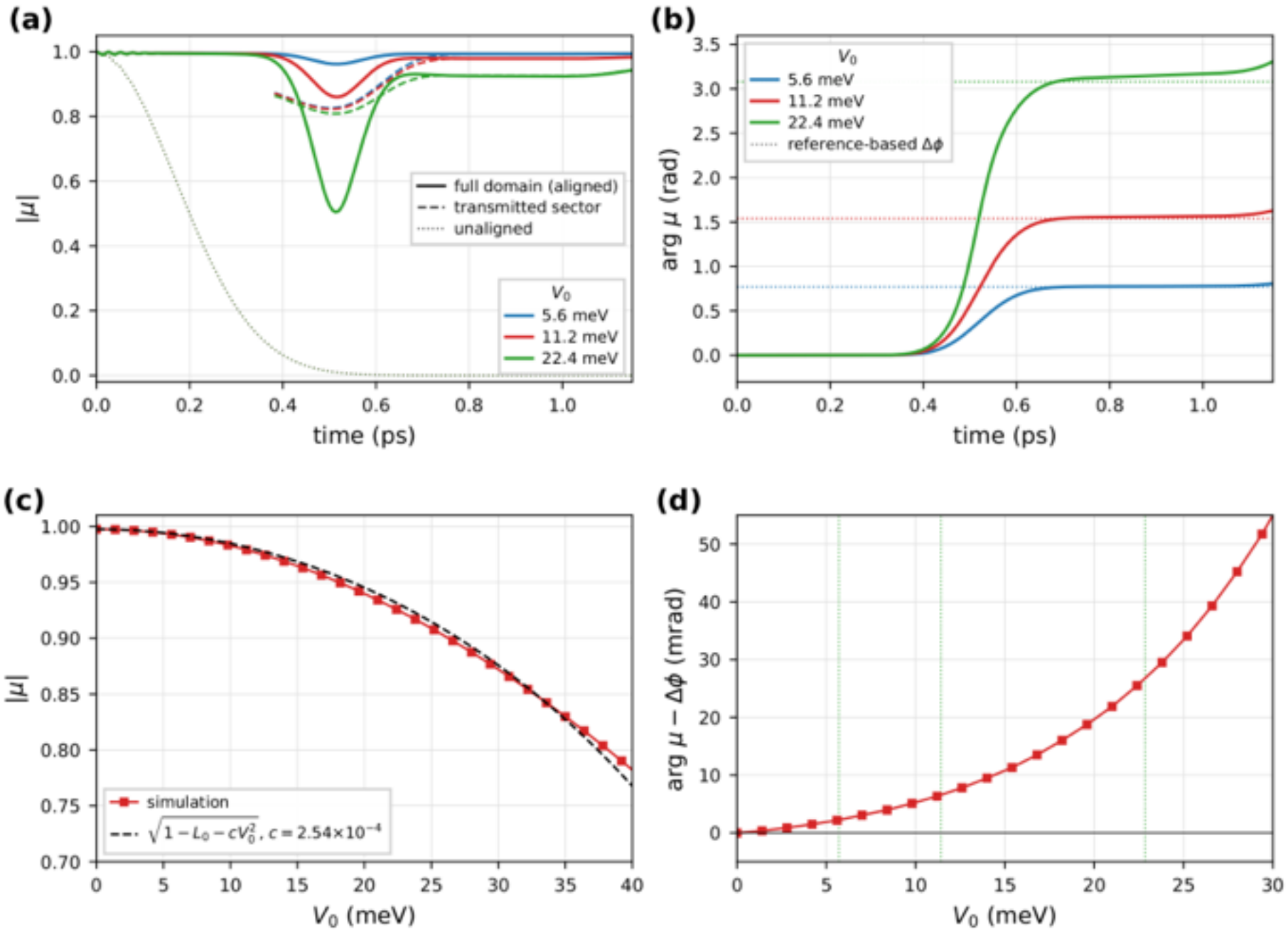

**Supplementary Fig. 2. Direct intervalley coherence.** **a**, $|\mu|(t)$ during transport for the three operating barrier heights (solid: full domain, aligned; dashed: transmitted sector; dotted: unaligned). **b**, $\arg\mu(t)$ forms entirely within the barrier-traversal window; dotted lines mark the reference-based $\Delta\phi$ at the same $V_0$. **c**, Visibility envelope $|\mu|(V_0)$ with the deflection-model law. **d**, Direct versus reference-based phase: $\arg\mu - \Delta\phi$ remains at the milliradian level across the operating window (dotted verticals: operating points).

# 4 Supplementary Note 4: Time-resolved probability current

This section presents a complementary transport-level diagnostic of the valley phase-control mechanism by tracking the expectation value of the probability current throughout the scattering process.

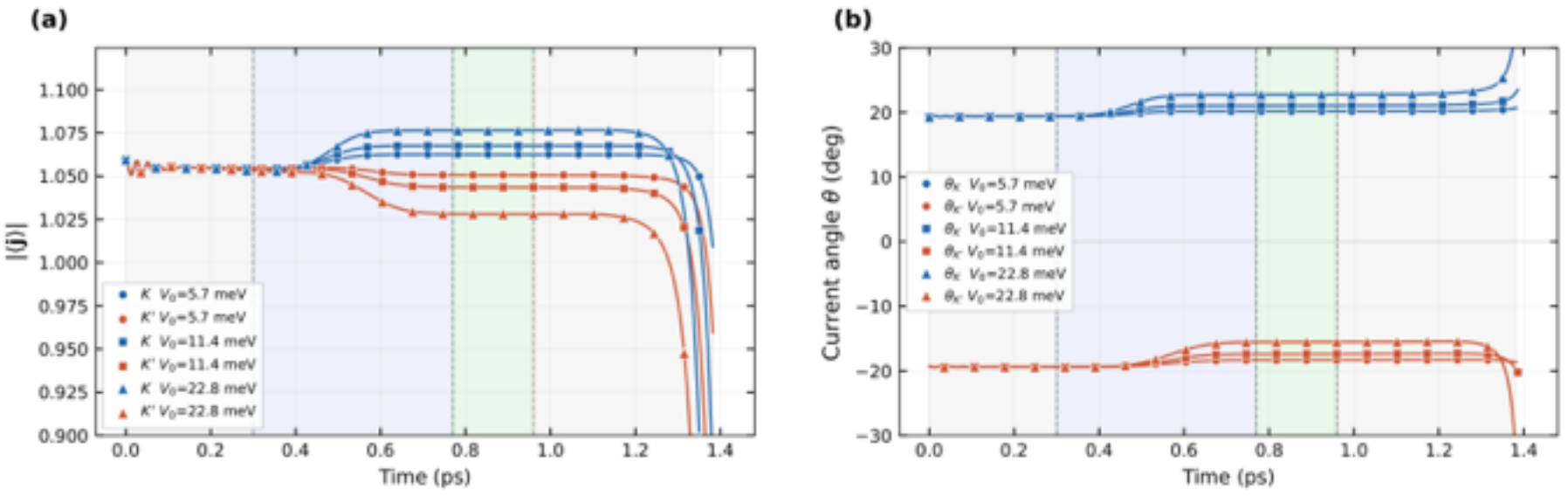

**Supplementary Fig. 3. Time-resolved probability-current magnitude and deflection angle during valley phase-control operation.** **a**, Normalized current magnitude $|\langle\mathbf{j}\rangle|$ as a function of time for both valleys ($K$, blue; $K'$, red) at three representative barrier heights: $V_0 = 5.7$ meV (circles), $11.4$ meV (squares), and $22.8$ meV (triangles). **b**, Valley-resolved current deflection angle $\theta$ versus time for the same operating points. In both panels, shaded regions delineate four temporal regimes: the white region (A) marks free propagation where $K$ and $K'$ currents are symmetric at $\theta_0 = \pm\arctan(\zeta_y) \approx \pm 19.3°$; the blue-shaded region (B) corresponds to active barrier interaction; the green-shaded region (C) identifies the postscattering plateau; and the gray-shaded region (D) captures the late-time drain absorption regime. The plateau values of the current deflection angle yield valley-angle splittings $\Delta\theta \approx 1.8°$, $3.6°$, and $7.2°$ at the three operating points, following the same $1:2:4$ ratio as the corresponding $\pi/4$, $\pi/2$, and $\pi$ phase targets.

Before the wave packet encounters the barrier (region A), the two valleys carry identical current magnitudes and opposite deflection angles $\theta_0 = \pm\arctan(\zeta_y)$, reflecting the bare tilt of the Dirac cone. For normal injection, this follows directly from the full tilted-Dirac current $\langle\mathbf{j}_\tau\rangle \propto (v_F, \tau w_y)$.

Upon entering the shaped barrier (region B), the tilt-induced asymmetry manifests as a progressive splitting: the $K$-valley current magnitude increases while $K'$ decreases, and their deflection angles diverge away from the symmetric initial value. In the postscattering plateau (region C), all observables converge to well-defined, barrier-height-dependent constants; critically, the valley-angle splitting $\Delta\theta = \theta_K - |\theta_{K'}|$ scales linearly with the applied barrier height, yielding 1.8°, 3.6°, and 7.2° at $V_0 = 5.7$, 11.4, and 22.8 meV. This 1:2:4 ratio directly mirrors the $\pi/4$:$\pi/2$:$\pi$ hierarchy of the overlap-extracted phase, making the valley-dependent current deflection a real-space signature of the differential dynamical phase accumulation. The current-magnitude asymmetry between valleys likewise grows with $V_0$, consistent with the onset of coherent deviation from the transported reference modes, which the main text identifies as the primary limitation on the ideal operating regime at elevated barrier heights. Within the plateau window, however, the magnitude splitting remains modest ($< 5\%$ at $V_0 = 22.8$ meV).

# 5 Supplementary Note 5: Tilted energy dispersion and candidate materials

The electrostatic phase-control mechanism under consideration fundamentally relies on a low-energy band structure that hosts two distinct valleys, separated in momentum space and characterized by Dirac cones with opposite tilts. The tilt parameter $\zeta = w_y/v_F$ sets the valley-dependent traversal phase: the relative phase $\Delta\phi \approx (V_0/\hbar v_F)\,\Delta D_{\mathrm{eff}}$ is controlled by the effective barrier-width difference $\Delta D_{\mathrm{eff}}$, which scales proportionally with $\zeta$, so a larger tilt yields a larger phase contrast per unit barrier height.

We now survey the material platforms in which quantitative values of the tilt parameter have been experimentally or theoretically reported: two-dimensional Dirac semimetals, three-dimensional Weyl and Dirac semimetals that can be accessed in exfoliated flake form, and organic charge-transfer salts.

***Two-dimensional Dirac semimetals.***

**8-*Pmmn* borophene ($\zeta = 0.37$).** Among tilted Dirac materials, the 8-*Pmmn* polymorph of monolayer borophene is distinguished by its precisely characterized tilt parameter. First-principles calculations attribute the tilt in 8-*Pmmn* borophene to the $p_z$ orbitals localized on the inner boron sublattice [5]. Subsequent tight-binding parameterizations yield characteristic velocities: $v_x = 0.86 \times 10^6$ m s$^{-1}$, $v_y = 0.69 \times 10^6$ m s$^{-1}$, and a tilt velocity $v_t = 0.32 \times 10^6$ m s$^{-1}$. The widely used tilt ratio, $\zeta = v_t/v_x \approx 0.37$, serves as a concrete parameter for modeling valley-dependent transport phenomena. The tilt vector in this material is oriented strictly along the armchair ($y$) direction and reverses sign between the two inequivalent valleys. This valley-contrasting tilt results in an oblique Klein tunneling angle of approximately 20.4° from normal incidence [6]. Atomic substitution, such as replacing selected boron sites with carbon atoms, enables continuous tuning of the tilt parameter. This method permits access to tilt values above and below the pristine 8-*Pmmn* value, providing a means to optimize device performance for specific valleytronic applications [7]. However, the 8-*Pmmn* phase has not yet been experimentally synthesized; all borophene samples grown on Ag(111) substrates correspond to structurally distinct polymorphs, such as the $\chi_3$, $\beta_{12}$, and 2-*Pmmn* phases [8, 9]. The predicted tilt of 0.37 nevertheless falls within a favorable operating range, the two-valley structure provides exactly the required two-level valley subspace, and the Dirac cone is well isolated from other bands over an energy window exceeding $\pm 0.5$ eV.

***Organic Dirac semimetals.***

**$\alpha$-(BEDT-TTF)$_2$I$_3$ ($\zeta \approx 0.8$).** The organic charge-transfer salt $\alpha$-(BEDT-TTF)$_2$I$_3$ under hydrostatic pressure ($\gtrsim 1.5$ GPa) is the first material in which a tilted Dirac cone was both predicted and experimentally confirmed [10, 11]. The low-energy Hamiltonian yields a tilt ratio of $\zeta \approx 0.8$ [12, 13], making it the most strongly tilted experimentally verified type-I Dirac material. The Fermi velocity is highly anisotropic ($v_{\max}/v_{\min} \approx 5$–$10$) and approximately two orders of magnitude below that of graphene ($\sim 10^4$ m s$^{-1}$) [14], which reduces the characteristic energy scales for gate operation accordingly. Although the large tilt ratio maximizes valley phase contrast per barrier, the low Fermi velocity poses challenges for device miniaturization, and the need for sustained hydrostatic pressure complicates integration into solid-state quantum circuits.

***Three-dimensional Weyl semimetals as few-layer flakes.***

In three-dimensional Weyl semimetals that are exfoliated into few-layer flakes, the in-plane component of the tilt vector assumes a role analogous to that in two-dimensional systems when considering two-terminal transport geometries. This analogy holds provided that the flake thickness remains larger than the characteristic interlayer coupling length scale, thereby preserving the bulk Weyl-node structure and, with it, the valley-dependent transport properties.

**TaAs-family monopnictides ($\zeta \sim 0.2$–$0.5$).** The transition metal monopnictides TaAs, TaP, NbAs, and NbP host 24 Weyl nodes each (12 W1 + 12 W2) and are prototypical type-I Weyl semimetals [15]. For TaAs specifically, the W2 tilt velocity components are $(v_{0x}, v_{0y}, v_{0z}) \approx (-0.85, +0.85, +1.40) \times 10^5$ m s$^{-1}$ against Fermi velocities of $(3.45, 2.65, 3.00) \times 10^5$ m s$^{-1}$ [16]. In few-layer flakes oriented with the $c$-axis normal to the substrate, the in-plane projection of the W2 tilt yields an effective two-dimensional tilt parameter of $\zeta_{\text{eff}} \approx 0.3$. These materials benefit from established crystal growth protocols and extensive experimental characterization. However, the presence of 24 Weyl nodes complicates the isolation of a clean two-valley subspace.

**$WTe_2$ and $MoTe_2$ ($\zeta > 1$; type-II).** Tungsten and molybdenum ditellurides are canonical type-II Weyl semimetals [17] with strongly over-tilted cones ($\zeta \approx 2.8$–$5.6$ for $WTe_2$) [18]. The type-II condition ($\zeta > 1$) fundamentally alters the Fermi surface topology, rendering the standard Klein tunneling mechanism inapplicable. Quantum confinement and substrate interactions in ultrathin flakes can modify the effective tilt, potentially reducing it below the critical value.

**$TaIrTe_4$ ($\zeta_{\text{eff}} \approx 0.37$ from experiment).** The ternary compound $TaIrTe_4$ is particularly instructive because its tilt has been extracted directly from experimental optical conductivity data [19]. Although first-principles calculations classify $TaIrTe_4$ as type-II [20], the observed interband optical response is best described by an effective type-I tilted Weyl model with $\zeta = 0.37$, which falls within a useful range for the present phase-element design.

**$NiTe_2$ and $PtTe_2$ ($\zeta > 1$ at DFT; sensitive to many-body corrections).** A significant recent finding is that many-body GW corrections can qualitatively alter the tilt classification: $PtTe_2$ changes from type-II to type-I under GW, while only $NiTe_2$ retains its type-II character [21]. This finding indicates that few-layer flakes of $PtTe_2$ may operate in the type-I regime relevant to the present phase-element design.

**$Co_3Sn_2S_2$ (magnetic Weyl semimetal).** The ferromagnetic kagomé compound $Co_3Sn_2S_2$ hosts tilted type-I Weyl nodes [22] and has been exfoliated into thin flakes [23], although the time-reversal symmetry breaking of its ferromagnetic order modifies the valley structure relevant to the present two-valley phase element.

**$YbMnBi_2$ (extreme anisotropy).** The canted antiferromagnet $YbMnBi_2$ exhibits a Fermi velocity anisotropy ratio exceeding $200{:}\,1$ [24], indicating that the tilt is confined almost entirely to a single momentum-space direction. This property may be advantageous if the transport axis is aligned with the tilt direction.

***Emerging predictions.*** Monolayer $Li_2N$ is predicted to realize a quasi-type-III Weyl state with $\zeta = 1.038$, which is close to the critical flat-band condition. Application of 3.7% tensile strain drives $\zeta$ to unity [25]. Gate-engineered graphene superlattices represent perhaps the most versatile platform, as the tilt can be tuned continuously from zero through the type-I/type-II transition purely by electrostatic means [26, 27].

***Summary.***

**Supplementary Table 2.** Candidate materials for electrostatic valley-phase control. The tilt parameter $\zeta = w_y/v_F$ governs the valley-dependent traversal phase across electrostatic barriers. Materials with $\zeta > 1$ are type-II and require modification of the barrier geometry or tilt tuning to access the type-I regime.

| Material | $\zeta$ | Type | Status | Dim. |
|---|---|---|---|---|
| Quinoid graphene | $\lesssim 0.05$ | I | Theor. | 2D |
| NbAs (W1) | $\approx 0.23$ | I | Expt. | 3D |
| $TaIrTe_4$ (expt. fit) | 0.37 | I* | Expt. | 3D |
| 8-$Pmmn$ borophene | 0.37 | I | Theor. | 2D |
| TaAs (W1) | $\approx 0.39$ | I | Expt. | 3D |
| TaAs (W2) | $\approx 0.47$ | I | Expt. | 3D |
| NbP (W2) | $\approx 0.50$ | I | Expt. | 3D |
| $\alpha$-$(BEDT\text{-}TTF)_2I_3$ | $\approx 0.8$ | I† | Expt. | 2D |
| $Li_2N$ monolayer | 1.04 | ~III | Theor. | 2D |
| $PtTe_2$ (GW-corrected) | $< 1$ ‡ | I | Expt. | 3D |
| $WTe_2$ | 2.8–5.6 | II | Expt. | 3D |

*Effective type-I tilt extracted from optical conductivity, despite DFT predicting type-II. †Near the type-I/type-II boundary; interlayer magnetoresistance data suggest $\zeta \to 1$. ‡DFT predicts type-II, but GW many-body corrections reverse the classification.

8-$Pmmn$ borophene ($\zeta = 0.37$) and $\alpha$-$(BEDT\text{-}TTF)_2I_3$ ($\zeta \approx 0.8$) serve as illustrative candidates because of their well-characterized tilt, distinct two-valley structure, and type-I Fermi-surface topology. In the context of three-dimensional Weyl semimetals available as few-layer flakes, $TaIrTe_4$ ($\zeta_{\mathrm{eff}} \approx 0.37$) and the W2 nodes of TaAs ($\zeta \approx 0.47$) provide representative effective tilt values. The proposed phase element developed here is most directly applicable to type-I systems featuring an isolated pair of transport valleys with opposite in-plane tilt. The simulations presented in the main text employ $\zeta_y = 0.35$, a value within the experimentally accessible range and representative of several effective tilt values inferred from experiments.

# 6 Supplementary Note 6: Robustness of electrostatic valley-phase control against static spatial disorder

To evaluate the experimental feasibility of electrostatic valley-phase control, we introduced static spatial disorder in the barrier region to model charge-puddle fluctuations arising from trapped charges in a realistic top-gate oxide. The disorder is parameterized by a root-mean-square amplitude $\delta V_{\mathrm{rms}} =$ 5% of the applied barrier height ($\approx$ 1.0 meV) and a spatial correlation length $L_c \approx$ 33 nm.

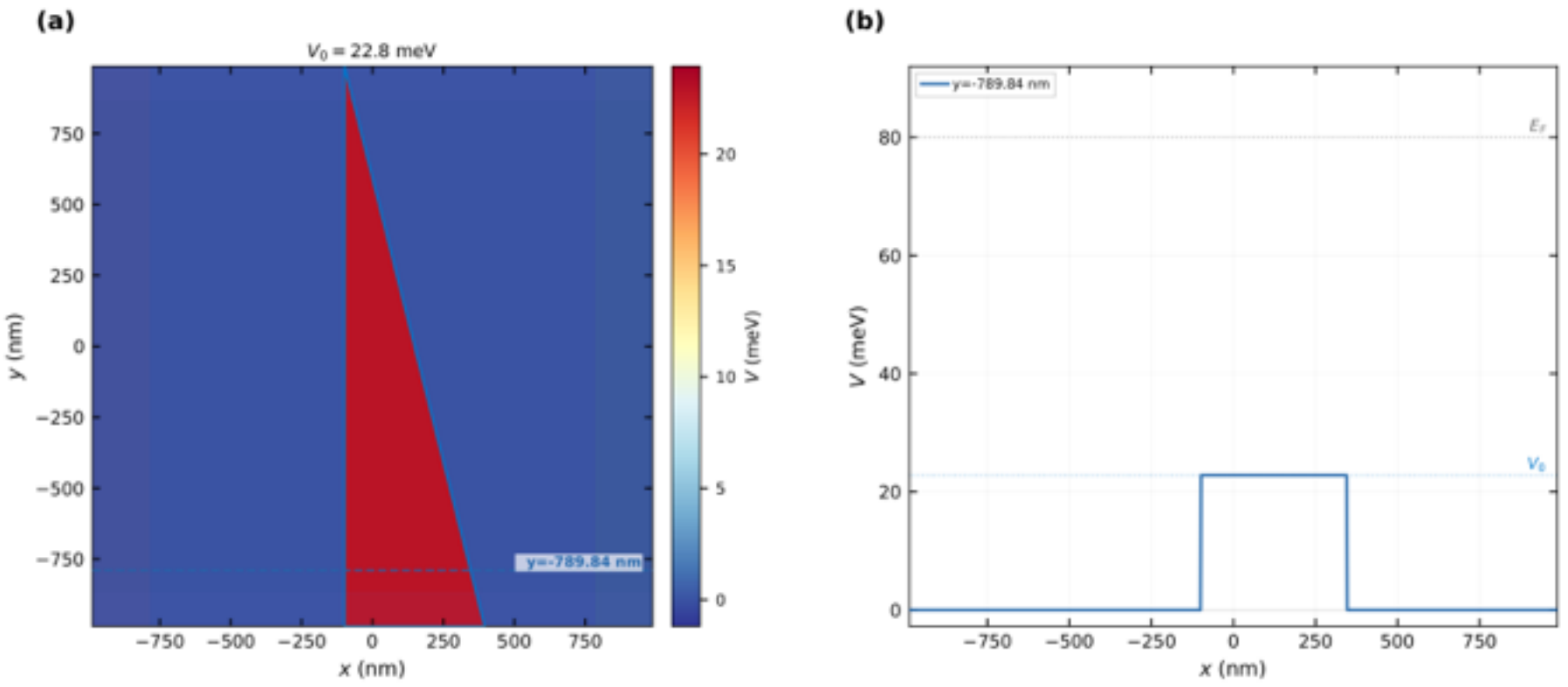

**Supplementary Fig. 4. Clean shaped-barrier potential landscape. a**, Two-dimensional electrostatic potential $V(x, y)$ of the shaped barrier at $V_0 = 22.8\,\text{meV}$. The barrier is defined by a vertical entrance interface and an oblique exit interface whose $x$-intercept increases linearly from the top to the bottom of the channel, producing the characteristic triangular geometry. The dashed line marks the transverse position of the line-out in panel (b). Light-shaded side bands indicate the source and drain absorbing regions; top and bottom bands denote the reflecting mass-wall confinement zones. **b**, Potential line-out $V(x)$ along the marked transverse coordinate, showing a clean rectangular cross-section with sharp interfaces at $V_0 = 22.8\,\text{meV}$. Horizontal dotted lines mark $E_{\text{F}} = 80\,\text{meV}$ and $V_0$.

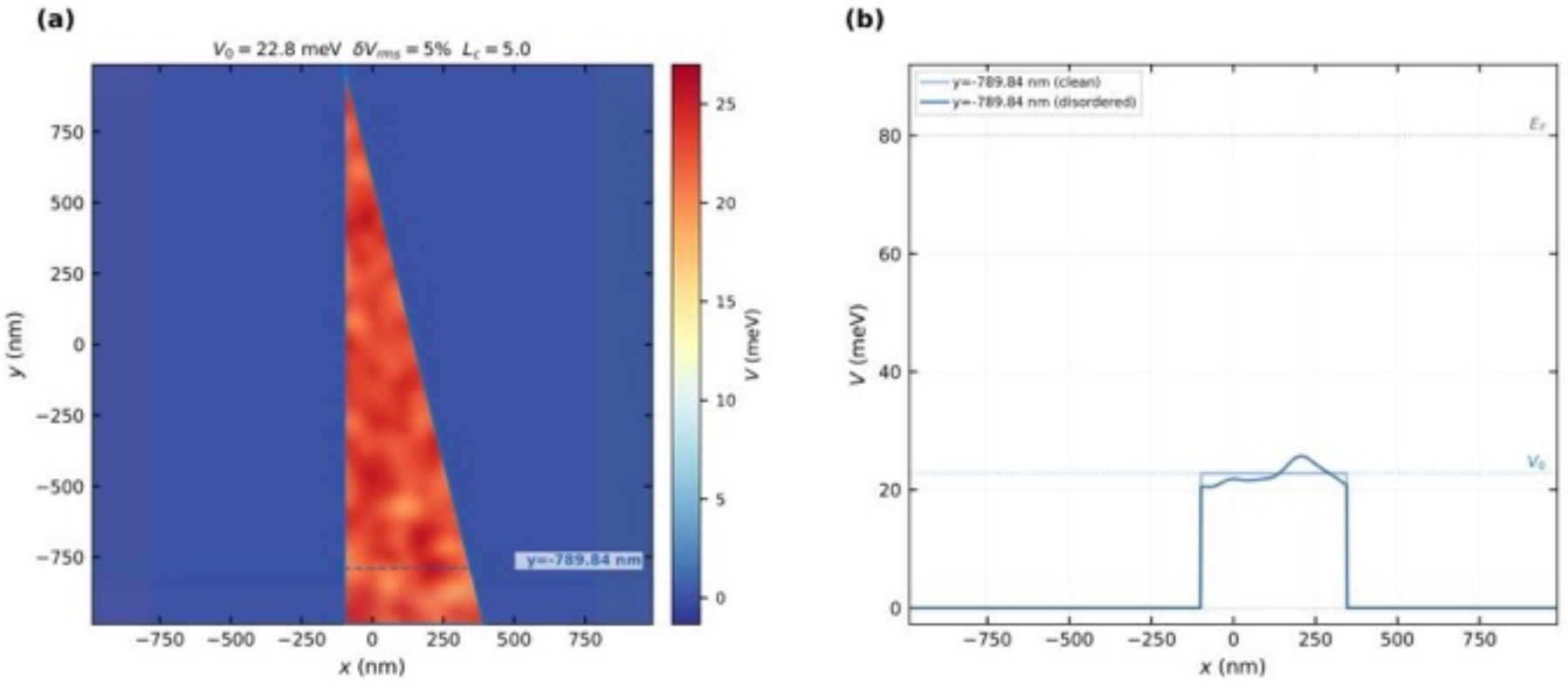

**Supplementary Fig. 5. Shaped-barrier potential landscape with static spatial disorder. a**, Two-dimensional potential $V(x, y)$ with charge-puddle disorder applied within the barrier region. The disorder is modeled as spatially correlated Gaussian fluctuations with $\delta V_{\text{rms}} = 5\%$ of $V_0$ ($\approx 1.0\,\text{meV}$) and correlation length $L_c \approx 33\,\text{nm}$, representative of oxide charge puddles in a realistic top-gate stack. The barrier boundaries remain identical to the clean case; the disorder modulates only the interior potential. **b**, Potential line-out comparing the clean profile (light blue) with the disordered realization (dark blue). The disorder introduces localized fluctuations on the order of $\pm 1\,\text{meV}$ about the mean barrier height while preserving the macroscopic rectangular envelope. These fluctuations are small relative to $E_{\text{F}}$ ($\delta V_{\text{rms}}/E_{\text{F}} \approx 1.3\%$), ensuring that the carrier remains in the single-mode propagation regime throughout.

## Quantitative comparison of clean and disordered gates

Supplementary Figs. 4 and 5 provide a comparative analysis of the resulting potential landscapes. The disorder modulates the barrier interior by $\pm 1\,\text{meV}$ while preserving the macroscopic geometry and, consequently, the valley-dependent effective widths. The transmitted wave-function profiles (shown in Supplementary Fig. 6) closely resemble the clean case in both centroid position and envelope shape.

The comprehensive barrier-height-sweep comparison (presented in Supplementary Fig. 7 and Supplementary Table 3) identifies two effects for the representative disorder realization examined in this study. First, the phase-versus-barrier-height slope increases by approximately 12% (from $0.0435\pi\,\text{meV}^{-1}$ to $0.0490\pi\,\text{meV}^{-1}$), indicating an increased effective scattering path as spatially correlated fluctuations alter the average propagation constant within the barrier. This change shifts the required operating barrier heights downward by approximately 9%, from 5.7, 11.4, and 22.8 meV to 5.2, 10.4, and 20.8 meV for the $\pi/4$, $\pi/2$, and $\pi$ operating points, respectively. This systematic offset can be compensated for by recalibrating the macroscopic top gate. Second, the mean overlap magnitude remains essentially unchanged relative to the clean case. At the $\pi/2$ and $\pi$ operating points, the disordered-barrier values are slightly larger than the clean-barrier values (by $+0.013$ and $+0.024$, respectively). This observation is consistent with the hypothesis that spatially correlated disorder softens the abrupt electrostatic edges, thereby enhancing Fermi-surface matching at the barrier interfaces. Throughout the low-barrier operating window, the valley polarization remains negligible ($|\eta| < 5 \times 10^{-4}$), and the transmission probabilities exceed 0.99. Because the disorder is smooth on the lattice scale, it does not activate the intervalley coupling channel; short-range intervalley perturbations are treated in Supplementary Note 1.

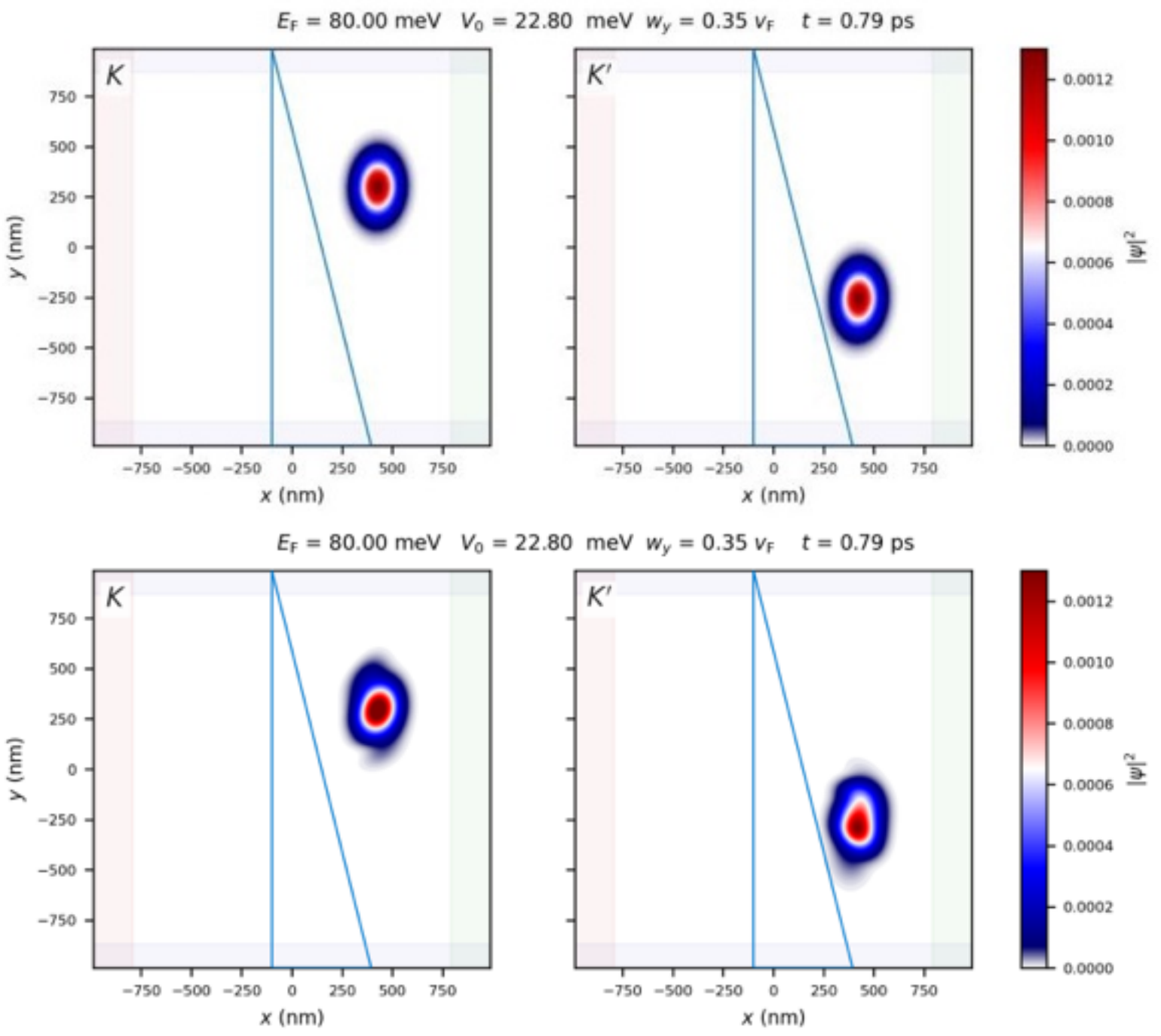

**Supplementary Fig. 6. Transmitted wave-function density for the clean (top) and disordered (bottom) barriers.** Probability density $|\psi|^2$ for the $K$ (left) and $K'$ (right) valleys after traversal at $V_0 = 22.8$ meV ($\pi$ operating point), evaluated at $t = 0.79$ ps within the postscattering plateau. The valleys emerge on opposite sides of the channel axis ($K$ at $y \approx +300$ nm, $K'$ at $y \approx -300$ nm), reflecting the tilt-induced transverse drift, and retain compact Gaussian envelopes with concentric rings under the 5% fluctuations: no fragmentation, scattering halos, or asymmetric distortion is apparent.

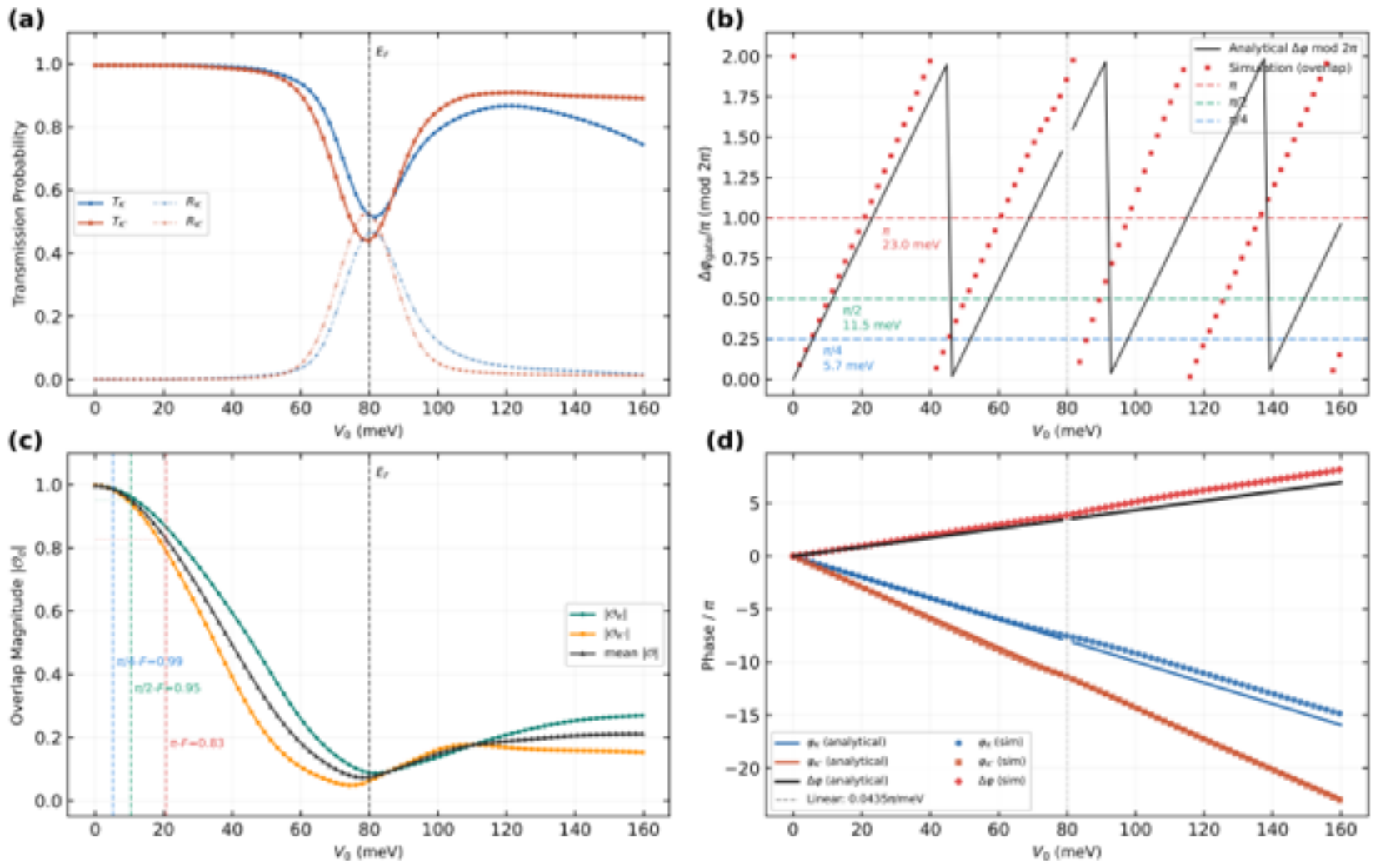

**Supplementary Fig. 7. Barrier-height dependence of transport, relative phase, and overlap quality under 5% static disorder.** Layout as in Fig. 2 of the main text, with $\delta V_{\mathrm{rms}} = 5\%$ charge-puddle disorder inside the barrier. **a**, Valley-resolved transmission and reflection: the low-barrier transmission remains near unity ($T > 0.99$) with negligible valley asymmetry. **b**, Relative phase $\Delta\phi/\pi$ (mod $2\pi$) with the analytical approximation; disorder steepens the slope from $0.0435\pi$ to $0.0490\pi\,\mathrm{meV}^{-1}$, shifting the operating points to 5.2, 10.4, and 20.8 meV. **c**, Overlap magnitudes $|\mathcal{O}_K|$, $|\mathcal{O}_{K'}|$ and their mean, $\overline{|\mathcal{O}|} = 0.99, 0.95, 0.83$ at the respective operating points—marginally above the clean case at $\pi/2$ and $\pi$. **d**, Unwrapped per-valley phases and their difference, with the steeper slope reflecting the increased effective scattering path.

**Supplementary Table 3.** Operating parameters and mean overlap magnitudes for the clean and disordered ($\delta V_{\mathrm{rms}} = 5\%$, $L_c \approx 33$ nm) barriers. For each target rotation, the operating barrier height is the $V_0$ at which the overlap-based relative phase crosses the target value. $\overline{|\mathcal{O}|}$ is the mean overlap magnitude at that operating point.

| | Clean barrier | | Disordered barrier | | |
|---|---|---|---|---|---|
| Operating point | $V_0$ (meV) | $\overline{\lvert\mathcal{O}\rvert}$ | $V_0$ (meV) | $\overline{\lvert\mathcal{O}\rvert}$ | $\Delta\overline{\lvert\mathcal{O}\rvert}$ |
| $\pi/4$ | 5.7 | 0.985 | 5.2 | 0.984 | −0.001 |
| $\pi/2$ | 11.4 | 0.946 | 10.4 | 0.960 | +0.013 |
| $\pi$ | 22.8 | 0.802 | 20.8 | 0.826 | +0.024 |
| Phase slope $\mathrm{d}(\Delta\phi)/\mathrm{d}V_0$ | | $0.0435\pi\,\mathrm{meV}^{-1}$ | | $0.0490\pi\,\mathrm{meV}^{-1}$ | (+12%) |
| Valley polarization $\lvert\eta\rvert$ at $\pi$ operating point | | $4.8 \times 10^{-4}$ | | $4.6 \times 10^{-4}$ | |

# 7 Supplementary Note 7: Split-operator Fourier propagation of the valley wave packet

This section provides the full specification of the numerical propagation scheme used to obtain all results reported in the main text. The method evolves the coupled four-component two-valley Dirac spinor on a uniform two-dimensional grid using a second-order split-operator Fourier algorithm.

***Hamiltonian.*** The spinors of the two valleys form a single four-component state $\Psi = (\psi_K, \psi_{K'})$ evolving under

$$H = \begin{pmatrix} H_K & U_{KK'} \\ U_{KK'}^{\dagger} & H_{K'} \end{pmatrix}, \tag{17}$$

$$H_\tau = \tau\hbar(w_x k_x + w_y k_y)\,\mathbb{I} + \hbar v_{\mathrm{F}}(\sigma_x k_x + \tau\sigma_y k_y) + V(x,y)\,\mathbb{I} + M(y)\,\sigma_z,$$

with valley index $\tau = \pm 1$ ($K$ and $K'$), where $\sigma_i$ are Pauli matrices in the pseudospin basis, $\mathbb{I}$ is the $2 \times 2$ identity, $V(x, y)$ is the electrostatic barrier, $M(y)$ is a transverse mass-confinement term, and $U_{KK'}(x, y)$ is an intervalley coupling field. The tilt enters as $\tau\hbar\,\mathbf{w} \cdot \mathbf{k}\,\mathbb{I}$—proportional to the identity in pseudospin space—so that it modifies the propagation direction of each valley without opening a gap or lifting the energy degeneracy; both the tilt reversal and the conjugate chirality of the $K'$ block ($\sigma_y \to -\sigma_y$) follow from time-reversal symmetry, $H_{K'}(\mathbf{k}) = H_K^*(-\mathbf{k})$ [15]. Because $V(x, y)$ and $M(y)$ are smooth on the lattice scale, they carry no Fourier weight at the valley separation $|K - K'|$ and cannot couple the valleys; $U_{KK'}$ represents atomic-scale perturbations, whose effect on the phase operation is quantified in Supplementary Note 1. All clean-device results correspond to the limit $U_{KK'} \to 0$, in which the coupled propagation reduces exactly to two independent valley evolutions.

***Mass-wall confinement.*** The reflecting side walls are implemented through a mass profile

$$M(y) = M_{\text{wall}} \max\left[\cos^2\left(\frac{\pi\, d_{\text{bot}}}{2W_y}\right),\ \cos^2\left(\frac{\pi\, d_{\text{top}}}{2W_y}\right)\right], \tag{18}$$

inside edge strips of width $W_y$, where $d_{\text{bot}} = y + L_y/2$ and $d_{\text{top}} = L_y/2 - y$, and $M(y) = 0$ in the channel interior. The $M(y)\,\sigma_z$ term opens a local gap at the channel edges, implementing the infinite-mass boundary condition for Dirac fermions [28, 29] in a smooth $\cos^2$ form that avoids spurious reflections at the inner edge of the wall region.

***Injected wave packet.*** The initial state is a normalized Gaussian wave packet,

$$\Psi_{\text{in}}(x, y) = \mathcal{N} \exp\left[-\frac{(x - x_0)^2}{4\sigma_x^2} - \frac{y^2}{4\sigma_y^2}\right] e^{i(k_{x0}\,x + k_{y0}\,y)}\ \chi_+(\mathbf{k}_0), \tag{19}$$

with the positive-energy eigenvector

$$\chi_+(\mathbf{k}_0) = \frac{1}{\sqrt{2}}\begin{pmatrix} 1 \\ e^{i\tau\phi_0} \end{pmatrix}, \qquad \phi_0 = \arg(k_{x0} + i k_{y0}). \tag{20}$$

Because the tilt term is proportional to the identity, it shifts the dispersion but does not change the eigenvector structure; the conjugate chirality of the $K'$ block enters only through the sign $\tau$ in the eigenvector phase. All results correspond to normal injection ($k_{y0} = 0$), so that $k_{x0} = k_0 = 0.8$ in simulation units (see Supplementary Note 8) and $\phi_0 = 0$.

***Split-operator time evolution.*** The Hamiltonian is decomposed into real-space and momentum-space parts:

$$H_r = \begin{pmatrix} V\,\mathbb{I} + M\sigma_z & U_{KK'} \\ U_{KK'}^\dagger & V\,\mathbb{I} + M\sigma_z \end{pmatrix}, \qquad H_k = \text{diag}(H_{k,K}, H_{k,K'}),$$
$$H_{k,\tau} = \tau\hbar(w_x k_x + w_y k_y)\,\mathbb{I} + \hbar v_{\text{F}}(\sigma_x k_x + \tau\sigma_y k_y). \tag{21}$$

One time step is then approximated by the second-order Strang splitting [30]:

$$\Psi(t + \Delta t) \approx e^{-iH_r\,\Delta t/2\hbar}\ \mathcal{F}^{-1}\ e^{-iH_k\,\Delta t/\hbar}\ \mathcal{F}\ e^{-iH_r\,\Delta t/2\hbar}\ \Psi(t), \tag{22}$$

with local error $O(\Delta t^3)$ [31]. For scalar coupling $U_{KK'} = u(x, y)\,\mathbb{I}$, the real-space half step factorizes into independent $2 \times 2$ blocks in the $(K_s, K'_s)$ subspace for each pseudospin component $s \in \{\uparrow, \downarrow\}$, each admitting the exact exponential

$$\exp\left[-i\begin{pmatrix} a & u \\ u^* & a \end{pmatrix}\frac{\Delta t}{2\hbar}\right] = e^{-ia\,\Delta t/2\hbar}\begin{pmatrix} \cos\theta & -i\,(u/|u|)\sin\theta \\ -i\,(u^*/|u|)\sin\theta & \cos\theta \end{pmatrix}, \tag{23}$$

$$\theta = \frac{|u|\,\Delta t}{2\hbar},$$

with $a = V + M$ for $s = \uparrow$ and $a = V - M$ for $s = \downarrow$; at $u = 0$ this reduces to the uncoupled phase factors $\psi_\uparrow \to e^{-i(V+M)\Delta t/2\hbar}\psi_\uparrow$, $\psi_\downarrow \to e^{-i(V-M)\Delta t/2\hbar}\psi_\downarrow$ in each valley. The momentum-space step factorizes into the two valley blocks and is evaluated analytically as a $2 \times 2$ unitary for each valley at every $(k_x, k_y)$ point [32]:

$$e^{-iH_{k,\tau}\Delta t/\hbar} = e^{-i\tau(\mathbf{w}\cdot\mathbf{k})\Delta t}\left[\cos(v_\mathrm{F} k\,\Delta t)\,\mathbb{I} - i\sin(v_\mathrm{F} k\,\Delta t)\,\frac{k_x\,\sigma_x + \tau\,k_y\,\sigma_y}{k}\right], \tag{24}$$

where $k = \sqrt{k_x^2 + k_y^2}$. In the absence of drains, the propagation is exactly unitary; non-unitarity enters only through the absorbing contact regions.

***Transmission and reflection extraction.*** The right and left contact absorbers are smooth $\cos^2$ masks:

$$m(d) = 1 - S\cos^2\left(\frac{\pi\,d}{2W_x}\right), \tag{25}$$

applied inside strips of width $W_x$ adjacent to each edge, with $m = 1$ in the interior. At each time step, the probability removed by the right absorber is added to the cumulative transmission $T_\tau(t)$ and that removed by the left absorber to the cumulative reflection $R_\tau(t)$. With reflecting $y$-boundaries, the probability budget satisfies

$$T_\tau(t) + R_\tau(t) + P_\tau^{\mathrm{rem}}(t) = 1, \tag{26}$$

where $P_\tau^{\mathrm{rem}}$ is the probability remaining on the grid. Runs are terminated automatically once $P_\tau^{\mathrm{rem}} < 3 \times 10^{-3}$.

***Probability current.*** For the tilted Dirac Hamiltonian, the probability current used in the real-space diagnostics (see Fig. 5 of the main text and Supplementary Fig. 3) contains both the pseudospin contribution and the tilt drift term:

$$\mathbf{j}_\tau(x, y, t) = \tau\,\mathbf{w}\,|\Psi_\tau|^2 + v_\mathrm{F}\,\Psi_\tau^\dagger\,(\sigma_x, \tau\sigma_y)\,\Psi_\tau, \tag{27}$$

where $|\Psi_\tau|^2 = |\psi_\uparrow|^2 + |\psi_\downarrow|^2$ and the operator carries the chirality of the corresponding valley block; all current-based diagnostics in this work use the operator matching the state's chirality. For the geometry used in the main text, $\mathbf{w} = (0, w_y)$, so the components are

$$j_x = 2v_\mathrm{F}\,\mathrm{Re}(\psi_\uparrow^*\,\psi_\downarrow), \qquad j_y = \tau w_y(|\psi_\uparrow|^2 + |\psi_\downarrow|^2) + 2\tau v_\mathrm{F}\,\mathrm{Im}(\psi_\uparrow^*\,\psi_\downarrow). \tag{28}$$

# 8 Supplementary Note 8: Analytical approximation of the phase model

The analytical curves shown alongside the numerical results in Fig. 2 of the main text serve as a single-mode approximation for comparison with the full wave-packet simulation. They are not used to define the reported relative-phase targets, which are extracted exclusively from the overlap method described in Supplementary Note 2. This section derives the analytical model from first principles.

***Dispersion and wavevectors.*** For a given device-frame incidence angle $\theta$ and signed tilt component $\tau w_y$, the outside-barrier wavevector of a carrier at energy $E_\mathrm{F}$ is

$$k_{\mathrm{F},\tau}(\theta) = \frac{E_\mathrm{F}}{\hbar\big(v_\mathrm{F} + \tau\,w_y\,\sin\theta\big)}, \qquad k_x = k_{\mathrm{F},\tau}\cos\theta, \qquad k_y = k_{\mathrm{F},\tau}\sin\theta. \tag{29}$$

Inside the barrier of height $V_0$, the longitudinal wavevector becomes

$$q_\tau = \frac{1}{\hbar v_{\rm F}} \sqrt{\left(E_{\rm F} - V_0 + \tau\, \hbar w_y\, k_y\right)^2 - \left(\hbar v_{\rm F}\, k_y\right)^2}, \tag{30}$$

with complex continuation in the evanescent regime. The corresponding barrier spinor angle is

$$\vartheta_\tau = \tan^{-1}\left(\frac{k_y}{q_\tau}\right), \tag{31}$$

and the sign $s = \mathrm{sgn}(E_{\rm F} - V_0)$ distinguishes electron-like and hole-like barrier states.

***Transmission amplitude.*** The complex transmission amplitude is the closed-form spinor-matching result for a rectangular tilted-Dirac barrier of width $D_\tau$:

$$t_\tau = \frac{e^{i(q_\tau - k_x)D_\tau}\,(1 + e^{2i\vartheta_\tau})\,(1 + e^{2i\theta})\, s}{\mathcal{D}_\tau}, \tag{32}$$

where the denominator is

$$\begin{aligned} \mathcal{D}_\tau \;&= e^{i(\vartheta_\tau+\theta)} - e^{2iq_\tau D_\tau + i(\vartheta_\tau+\theta)} + s + s\, e^{2iq_\tau D_\tau + 2i\vartheta_\tau} \\ &+ s\, e^{2iq_\tau D_\tau + 2i\theta} + s\, e^{2i(\vartheta_\tau+\theta)} + s^2\, e^{i(\vartheta_\tau+\theta)} - s^2\, e^{2iq_\tau D_\tau + i(\vartheta_\tau+\theta)}. \end{aligned} \tag{33}$$

This expression is the standard transfer-matrix result for tilted-Dirac electron optics, well established in the literature for oblique Klein tunneling [6, 33].

***Analytical valley phases.*** The per-valley analytical phases are defined as

$$\phi_\tau^{\rm an}(V_0) = \arg t_\tau(E_{\rm F}, V_0, D_\tau, \theta), \tag{34}$$

and the analytical relative phase is

$$\Delta\phi^{\rm an}(V_0) = \phi_K^{\rm an}(V_0) - \phi_{K'}^{\rm an}(V_0). \tag{35}$$

For all figures in the main text, the analytical approximation is evaluated at normal injection ($\theta = 0$), using the same signed tilt reversal $w_y \to \pm w_y$ as in the numerical simulations.

***Effective barrier width from shaped geometry.*** The only geometric input to the analytical model is the valley-dependent effective width $D_\tau$. This is derived from the semiclassical group velocity and the barrier geometry as follows.

For normal injection ($k_y = 0$) and purely transverse tilt ($w_x = 0$), the outside-barrier group velocity is

$$\mathbf{v}_{g,\tau} = \nabla_{\mathbf{k}}\, E_\tau = \tau\, \mathbf{w} + v_{\rm F}\, \hat{\mathbf{x}}, \tag{36}$$

giving $v_x = v_{\rm F}$ and $v_{y,\tau} = \tau\, w_y$. The wave-packet centroid, therefore, follows a straight valley-dependent ray:

$$y_\tau(x) = \tau\, \frac{w_y}{v_{\rm F}}\, (x + d), \tag{37}$$

where $d$ is the source-to-entrance-interface distance.

The oblique exit interface of the shaped barrier is parameterized as

$$x_R(y) = W\left(\frac{L_y/2 - y}{L_y}\right), \tag{38}$$

where $W$ is the maximum barrier width. Solving Eq. (37) and the exit-interface equation for their intersection yields the effective traversal length:

$$\boxed{D_\tau = W\,\frac{L_y/2 - \tau\,\zeta_y\,d}{L_y + \tau\,\zeta_y\,W}}, \tag{39}$$

where $\zeta_y = w_y/v_\mathrm{F}$ is the tilt parameter. The difference

$$\Delta D_\mathrm{eff} = D_{K'} - D_K \tag{40}$$

is the valley-dependent effective barrier width difference that appears in the relative-phase formula (Eq. 9 of the main text). Writing $\tan\alpha = W/L_y$ for the exit-interface angle, it takes the closed form

$$\Delta D_\mathrm{eff} = \frac{\zeta_y\,W L_y\,(W + 2d)}{L_y^2 - \zeta_y^2 W^2} = \frac{\zeta_y\,\tan\alpha\,(W + 2d)}{1 - \zeta_y^2 \tan^2\alpha}, \tag{41}$$

quoted in Appendix D of the main text: linear in the tilt for $\zeta_y\tan\alpha \ll 1$, with the geometric enhancement factor $(1 - \zeta_y^2\tan^2\alpha)^{-1}$, and linear in the interface angle at small $\alpha$. Supplementary Fig. 8 compares the predicted phase slope $\mathrm{d}(\Delta\phi)/\mathrm{d}V_0 = \Delta D_\mathrm{eff}/\hbar v_\mathrm{F}$ with simulations over four tilts ($\zeta_y = 0.15$–$0.45$) and four interface angles ($\alpha = 8.5°$–$16.7°$): all combinations agree to within +0.5–0.7%—a small, uniform finite-wavelength correction above the semiclassical chord—across a factor-8 range of slopes.

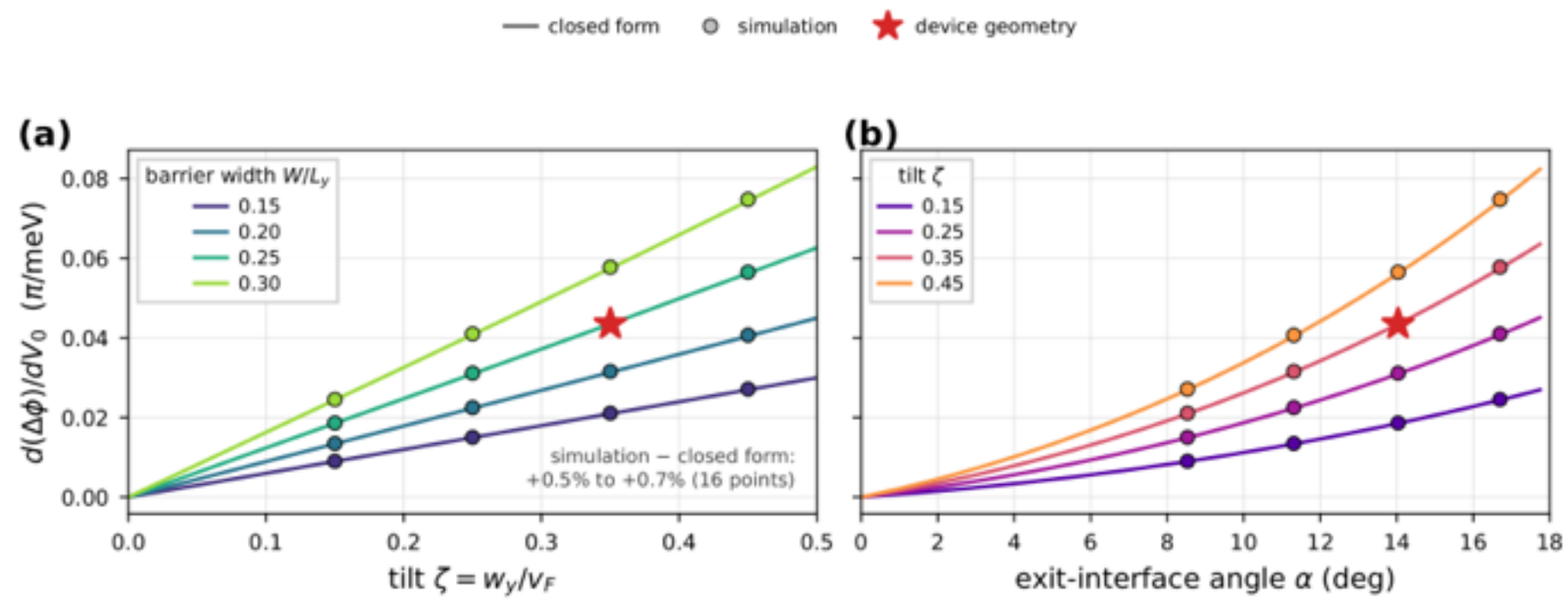

**Supplementary Fig. 8. Closed-form phase slope versus tilt and interface angle. a**, Phase slope $\mathrm{d}(\Delta\phi)/\mathrm{d}V_0$ versus tilt $\zeta_y$ for four barrier widths $W/L_y$; lines: Eq. (41), points: simulation. The star marks the device geometry. **b**, The same data versus the exit-interface angle $\alpha$ for four tilts.

Two scales bound the validity of Eq. (41). The geometric divergence at $\zeta_y\tan\alpha \to 1$ corresponds to the exit ray skimming the oblique interface; for the present geometry, this would require $\zeta_y = 4$, far outside the physical regime. The physical boundary is the type-I/type-II transition at $\zeta_y \to 1$, where the Fermi surface opens, single positive-energy injection and the two-interface refraction analysis fail, and the low-barrier expansion retains validity only for decreasing $V_0/E_\mathrm{F}$ (cf. Supplementary Note 5).

***Low-barrier linear approximation.*** In the low-barrier regime ($V_0 \ll E_\mathrm{F}$), the barrier does not significantly modify the carrier trajectory, and the transmission phase through a rectangular barrier of width $D_\tau$ reduces to $\phi_\tau \approx -(V_0/\hbar v_\mathrm{F})\,D_\tau$. The relative valley phase is then

$$\boxed{\Delta\phi \approx \frac{V_0}{\hbar v_\mathrm{F}}\,\Delta D_\mathrm{eff}}, \tag{42}$$

which is Eq. (9) of the main text. This gives the linear slope

$$\frac{\mathrm{d}(\Delta\phi)}{\mathrm{d}V_0} = \frac{\Delta D_\mathrm{eff}}{\hbar v_\mathrm{F}}. \tag{43}$$

For the device parameters ($\zeta_y = 0.35$, $W = 75$ sim. units, $d = 40$ sim. units, $L_y = 300$ sim. units, $a_0 = 6.58\,\text{nm/sim. unit}$), Eq. (39) gives $D_K \approx 31.26$ and $D_{K'} \approx 44.93$ sim. units, so $\Delta D_{\text{eff}} \approx 13.67$ sim. units $\approx 89.9\,\text{nm}$, yielding a phase slope of $0.0435\pi\,\text{meV}^{-1}$—in excellent agreement with the numerical result extracted from the simulation data (see Fig. 2d of the main text).

The $\pi$-rotation barrier height is

$$V_\pi = \frac{\pi\,\hbar v_\text{F}}{\Delta D_{\text{eff}}} \approx 23\,\text{meV}, \tag{44}$$

and the standard operating barrier heights follow the $1{:}\,2{:}\,4$ ratio:

$$V_{\pi/4} \approx 5.7\,\text{meV}, \qquad V_{\pi/2} \approx 11.4\,\text{meV}, \qquad V_\pi \approx 22.8\,\text{meV}. \tag{45}$$

***Thermal smearing of the extracted phase.*** The reference-subtracted phase is energy-independent at leading order: at normal incidence the analytic result $\Delta\phi = V_0\,\Delta D_{\text{eff}}/\hbar v_\text{F}$ contains no explicit carrier energy, because massless carriers traverse the barrier at $v_\text{F}$ regardless of $E$. Thermal smearing, therefore, enters only through the energy dependence of the refraction corrections, which are of relative order $V_0/E_\text{F}$. Expanding the phase integral about $E_\text{F}$ over the thermal window $k_\text{B}T$ gives

$$\delta(\Delta\phi) \;\lesssim\; \Delta\phi\,\frac{V_0}{E_\text{F}}\,\frac{k_\text{B}T}{E_\text{F}}, \tag{46}$$

i.e., $\delta(\Delta\phi) \approx 4\,\text{mrad}$ at $T = 4\,\text{K}$ at the $\pi$ operating point ($k_\text{B}T = 0.34\,\text{meV}$), far below the phase targets. Beyond the standard condition $k_\text{B}T \ll E_\text{F}$ of low-temperature coherent transport, the phase element itself therefore imposes no additional temperature constraint.